\documentclass[12pt]{article}
\usepackage{latexsym}
\usepackage{amsmath,amsfonts}
\usepackage{times}
\allowdisplaybreaks[4]

\catcode`\@=11

\newcount\hour
\newcount\minute
\newtoks\amorpm \hour=\time\divide\hour by 60\minute
=\time{\multiply\hour by 60 \global\advance\minute by-\hour}
\edef\standardtime{{\ifnum\hour<12 \global\amorpm={am}%
        \else\global\amorpm={pm}\advance\hour by-12 \fi
        \ifnum\hour=0 \hour=12 \fi
        \number\hour:\ifnum\minute<10
        0\fi\number\minute\the\amorpm}}
\edef\militarytime{\number\hour:\ifnum\minute<10
0\fi\number\minute}

\def\draftlabel#1{{\@bsphack\if@filesw {\let\thepage\relax
   \xdef\@gtempa{\write\@auxout{\string
      \newlabel{#1}{{\@currentlabel}{\thepage}}}}}\@gtempa
   \if@nobreak \ifvmode\nobreak\fi\fi\fi\@esphack}
        \gdef\@eqnlabel{#1}}
\def\@eqnlabel{}
\def\@vacuum{}
\def\marginnote#1{}
\def\draftmarginnote#1{\marginpar{\raggedright\scriptsize\tt#1}}
\def\draft{
        \pagestyle{plain}
        \overfullrule=2pt
        \oddsidemargin -.5truein
        \def\@oddhead{\sl \phantom{\today\quad\militarytime} \hfil
        \smash{\Large\sl DRAFT} \hfil \today\quad\militarytime}
        \let\@evenhead\@oddhead
        \let\label=\draftlabel
        \let\marginnote=\draftmarginnote
        \def\ps@empty{\let\@mkboth\@gobbletwo
        \def\@oddfoot{\hfil \smash{\Large\sl DRAFT} \hfil}
        \let\@evenfoot\@oddhead}
        \def\@eqnnum{(\theequation)\rlap{\kern\marginparsep\tt\@eqnlabel}%
        \global\let\@eqnlabel\@vacuum}  }

\newcommand{\rf}[1]{(\ref{#1})}
\renewcommand{\theequation}{\thesection.\arabic{equation}}
\renewcommand{\thefootnote}{\fnsymbol{footnote}}
\newcommand{\newsection}{    
\setcounter{equation}{0}\section}

\def\appendix#1{\addtocounter{section}{1}\setcounter{equation}{0}
\renewcommand{\thesection}{\Alph{section}}
\section*{Appendix \thesection\protect\indent \parbox[t]{11.15cm}{#1}}
\addcontentsline{toc}{section}{Appendix \thesection\ \ \ #1}}

\def\asf{{\sf a}}
\def\bsf{{\sf b}}
\def\csf{{\sf c}}
\def\esf{{\sf e}}
\def\Nsf{{\sf N}}

\def\nline{\,\nabla\kern -0.7em\raise0.2ex\hbox{/}\,\,}
\def\yline{\,y\kern -0.47em /}
\def\aline{\,a\kern -0.49em /}
\def\parline{\,\partial\kern -0.55em /\,\,}

\newcommand{\Eo}{\mathbb{E}}

\newcommand{\Mo}{\mathbb{M}}
\newcommand{\No}{\mathbb{N}}
\newcommand{\Po}{\mathbb{P}}

\newcommand{\Zo}{\mathbb{Z}}

\def\be{\begin{equation}}
\def\ee{\end{equation}}
\def\beq{\begin{eqnarray}}
\def\eeq{\end{eqnarray}}

\def\Rsm{{\scriptscriptstyle R}}
\def\Lsm{{\scriptscriptstyle L}}

\def\smpt{{\scriptscriptstyle [2]}}
\def\smp3{{\scriptscriptstyle [3]}}

\def\smpn{{\scriptscriptstyle [n]}}

\def\Jbf{{\bf J}}
\def\Ebf{{\bf E}}

\def\Mbf{{\bf M}}
\def\Pbf{{\bf P}}

\def\Xbf{{\bf X}}
\def\Vbf{{\bf V}}

\def\ibf{{\bf i}}
\def\iibf{{\bf ii}}
\def\iiibf{{\bf iii}}
\def\ivbf{{\bf iv}}
\def\vbf{{\bf v}}
\def\vibf{{\bf vi}}

\def\NN{{\cal N}}
\def\PP{{\cal P}}

\def\Nsf{{\sf N}}

\def\half{\frac{1}{2}}

\def\Cb{{\bar{C}}}

\def\Vb{{\bar{V}}}
\def\vb{{\bar{v}}}

\def\irm{{\rm i}}

\def\dyn{{\rm dyn}}
\def\minrm{{\rm min}}
\def\maxrm{{\rm max}}

\def\diff{{\rm diff}}
\def\GP{{\rm GP}}

\def\ext{{\rm ext}}

\def\betach{\check{\beta}}

\begin{document}

\begin{flushright}
FIAN-TD-2019-18 \ \ \ \ \ \\
arXiv: 1909.05241V2
\end{flushright}

\vspace{1cm}

\begin{center}

{\Large \bf Cubic interactions for arbitrary spin \hbox{$\NN$}-extended

\medskip

massless supermultiplets in 4d flat space}

\vspace{2.5cm}

R.R. Metsaev\footnote{ E-mail: metsaev@lpi.ru }

\vspace{1cm}

{\it Department of Theoretical Physics, P.N. Lebedev Physical
Institute, \\ Leninsky prospect 53,  Moscow 119991, Russia }

\vspace{3cm}

{\bf Abstract}

\end{center}

$\NN$-extended massless arbitrary integer and half-integer spin supermultiplets in four dimensional flat space are studied in the framework of light-cone gauge formalism.
For such multiplets, by using light-cone momentum superspace,
we build unconstrained light-cone gauge superfield formulation. The superfield formulation is used to develop a superspace representation for all cubic interactions vertices of the $\NN$-extended massless supermultiplets.
Our suitable treatment of the light-cone gauge superfields allows us to obtain attractively simple superspace representation for the cubic interaction vertices.
Superspace realization of relativistic symmetries of the $\NN$-extended Poincar\'e superalgebra on space of interacting fields is also obtained.

\vspace{3cm}

Keywords: N-extended supersymmetric higher-spin fields, light-cone gauge formalism, interaction vertices.

\newpage
\renewcommand{\thefootnote}{\arabic{footnote}}
\setcounter{footnote}{0}

\section{ \large Introduction}

In view of aesthetic features, $\NN$-extended supersymmetric theories have attracted considerable interest during long period of time. As is known, light-cone gauge approach
offers considerable simplifications for study of supersymmetric theories. For this reason, the $\NN$-extended supersymmetric theories have extensively been studied in the framework of this approach. We mention application of light-cone formalism for the investigation of ultraviolet finiteness of $\NN=4$ supersymmetric YM theory in Refs.\cite{Brink:1982wv,Mandelstam:1982cb}. Also we note that the light-cone gauge formulation of type $IIB$ supergravity theories in $10d$ flat space and $AdS_5 \times S^5$ space was developed in the respective Ref.\cite{Green:1982tk} and Ref.\cite{Metsaev:1999gz}, while the study of type $IIA$ and $IIB$ light-cone gauge superstring field theories may be found in Refs.\cite{Green:1983hw}. Recent interesting application of light-cone formalism for studying $\NN=8$ supergravity may be found in Refs.\cite{Kallosh:2009db,Ananth:2017xpj}.

Attractive example of application of light-cone gauge formalism is a supersymmetric higher-spin massless field theory. This is to say that, in the framework of light-cone gauge approach, a cubic interaction vertex of the {\it scalar} $\NN$-extended massless supermultiplet with arbitrary $\NN=4\No$ in $4d$ flat space was obtained in Ref.\cite{Bengtsson:1983pg}, while in Ref.\cite{Metsaev:2019dqt}, for the case of {\it arbitrary spin} (integer and half-integer) $\NN=1$ massless supermultiplets in $4d$ flat space, we obtained the full list of cubic interaction vertices. Result in Ref.\cite{Metsaev:2019dqt} provides the $\NN=1$ supersymmetric completion for all cubic interaction vertices for  arbitrary spin bosonic massless fields obtained in Refs.\cite{Bengtsson:1983pd,Bengtsson:1986kh}.%
\footnote{ In the recent time, the study of cubic interactions of higher-spin $\NN=1$ massless supermultiplets by using gauge invariant supercurrents, may be found in Refs.\cite{Buchbinder:2017nuc}-\cite{Gates:2019cnl}.}
In this paper, we consider {\it arbitrary spin} (integer and half-integer) $\NN$-extended massless supermultiplets with arbitrary $\NN=4\No$ in the $4d$ flat space.
For such supermultiplets, our aim is to find all cubic interaction vertices. To this end, as in Refs.\cite{Metsaev:2019dqt}, we prefer to use a light-cone gauge unconstrained superfields that are defined in a light-cone momentum superspace. We note that, in the past, the light-cone momentum superspace has fruitfully been used in many important and interesting studies of supergravity and superstring theories.
As example of attractive use of the momentum superspace we mention the building of $IIB$ supergravity in $10d$ flat space and superstring field theories in $10d$ flat space in the respective Ref.\cite{Green:1982tk} and Ref.\cite{Green:1983hw}. The momentum superspace turns also out to be very convenient for studying  supergravity in $11d$  flat space \cite{Metsaev:2004wv} and $IIB$ supergravity in $AdS_5\times S^5$ space \cite{Metsaev:1999gz}. In this paper, using Grassmann momentum entering the light-cone momentum superspace, we collect fields of $\NN$-extended massless supermultiplets into a suitable unconstrained light-cone gauge superfields and use such superfields to construct a full list of cubic interaction vertices. We note that it is  the formalism of unconstrained light-cone gauge superfields that provides us a possibility to build attractively simple expressions for cubic vertices and allows us to obtain the full classification of cubic interactions.

Some long term motivations for our study of supersymmetric higher-spin field theory which are beyond the scope of this paper may be found in Conclusions.

Our paper is organized in the following way.

In Sec.\ref{sec-02}, we start with brief review of light-cone coordinates frame and discuss general structure of the $\NN$-extended Poincar\'e superalgebra. We discuss a field content that enters arbitrary spin (integer and half-integer) massless $\NN$-extended supermultiplets. After that, we introduce our $\NN$-extended momentum superspace and provide the explicit description of light-cone gauge unconstrained superfields which are defined on such superspace.

Section \ref{sec-03} is devoted to description of general structure of $n$-point interaction vertices for theories of interacting fields. We provide a detailed description of constraints that are imposed by kinematical symmetries of the $\NN$-extended Poincar\'e superalgebra on $n$-point interaction vertices.

In Sec.\ref{sec-04}, we restrict out attention to cubic vertices. First, we adopt general kinematical constraints of the $\NN$-extended Poincar\'e superalgebra obtained in Sec.\ref{sec-03} to the case of cubic vertices. Second, we derive constraints imposed on the cubic vertices by  dynamical symmetries of the  $\NN$-extended Poincar\'e superalgebra. Third, we formulate light-cone gauge dynamical principle and, finally, we  present the complete system of equations that allows us to fix the cubic vertices unambiguously.

In Sec.\ref{sec-05}, we present our main result in this paper. We show explicit expressions for all cubic vertices that describe interactions of arbitrary spin $\NN$-extended massless supermultiplets. We start with the presentation of the superspace form of the cubic interaction vertices. After that, we present the restrictions on allowed values of $\NN$ and superfields helicities entering our cubic interaction vertices. These restrictions provide the classification of the cubic vertices that can be build for arbitrary spin $\NN$-extended supermultiplets in the framework of light-cone gauge approach. Also we discuss representation of the cubic vertices in terms of the component fields.

Sec.\ref{concl} summarizes our conclusions.

In Appendix A, we present our notation and conventions. In Appendix B, we describe properties of our light-cone gauge superfields.
In appendix C, we outline the derivation of the superspace cubic interaction vertices.

\newsection{ \large Light-cone gauge superfield formulation of free  $\NN$-extended  massless supermultiplets }\label{sec-02}

\noindent {\bf Light-cone coordinates frame}. We consider light-cone gauge fields by using a helicity basis. Therefore we start with the description of light-cone coordinates frame. In the flat space $R^{3,1}$, the Lorentz basis coordinates are denoted as $x^\mu$, $\mu=0,1,2,3$, while the light-cone basis coordinates denoted as $x^\pm$, $x^\Rsm$, $x^\Lsm$ are expressed in terms of $x^\mu$ as

\be \label{02092019-man-01}
x^\pm \equiv \frac{1}{\sqrt{2}}(x^3  \pm x^0)\,,\qquad
\qquad x^\Rsm \equiv \frac{1}{\sqrt{2}}(x^1 + \irm x^2)\,,\qquad x^\Lsm \equiv \frac{1}{\sqrt{2}}(x^1 - \irm x^2)\,.
\ee
Throughout this paper, the coordinate $x^+$ is taken to be a time-evolution parameter. Let $X^\mu$ be a vector of the $so(3,1)$ Lorentz algebra. In the light-cone basis \rf{02092019-man-01} the  $X^\mu$, is decomposed as $X^+,X^-,X^\Rsm$, $X^\Lsm$. Using notation $\eta_{\mu\nu}$ for the mostly positive flat metric of $R^{3,1}$, we note that a scalar product of the $so(3,1)$ Lorentz algebra vectors $X^\mu$ and $Y^\mu$ is represented in the following way:
\be \label{02092019-man-02}
\eta_{\mu\nu}X^\mu Y^\nu = X^+Y^- + X^-Y^+ + X^\Rsm Y^\Lsm + X^\Lsm Y^\Rsm\,.
\ee
Relation \rf{02092019-man-02} implies that, in the light-cone basis, non-vanishing elements of the $\eta_{\mu\nu}$ are given by $\eta_{+-} = \eta_{-+}=1$, $\eta_{\Rsm\Lsm} = \eta_{\Lsm\Rsm} = 1$. We note then that the covariant and contravariant vectors $X_\mu$, $X^\mu$ are related as follows: $X^+=X_-$, $X^-=X_+$, $X^\Rsm=X_\Lsm$, $X^\Lsm=X_\Rsm$.

\noindent {\bf Extended Poincar\'e superalgebra in light-cone frame}. The method proposed in Ref.\cite{Dirac:1949cp} reduces the problem of finding a light-cone gauge dynamical system to the problem of finding a solution of commutation relations for algebra of basic symmetries. For field theories with extended supersymmetries in flat space, the basic symmetries are governed by the extended Poincar\'e superalgebra. Therefore in order to fix our notation we now discuss a general structure of the extended Poincar\'e superalgebra.

For the case of the $R^{3,1}$ space, the $\NN$-extended Poincar\'e superalgebra consists  the translation generators $P^\mu$, the generators of the $so(3,1)$ Lorentz algebra $J^{\mu\nu}$, Majorana supercharges $Q^{\alpha i}$, $Q_i^\alpha$, and $su(\NN)$ R-symmetry algebra generators $J^i{}_j$. Explicit light-cone form of commutation relations of the extended Poincar\'e superalgebra we use in this paper may be found in Appendix A. Here we note that, in light-cone basis \rf{02092019-man-01}, generators of the extended Poincar\'e superalgebra can be separated into the following two groups:

{\small
\beq
\label{02092019-man-03} && \hspace{-1.6cm}
P^+, \ \
P^\Rsm,\ \ \
P^\Lsm, \ \ \ \
J^{+\Rsm},\ \ \ \
J^{+\Lsm},\ \
J^{+-},\ \
J^{\Rsm\Lsm},\ \
Q_i^{+\Rsm},\ \
Q^{+\Lsm i},\ \
J^i{}_j,
\hspace{0.5cm}\hbox{ kinematical generators};
\\
\label{02092019-man-04}  && \hspace{-1.6cm}
P^-, \ \
J^{-\Rsm}, \ \
J^{-\Lsm}, \ \
Q^{-\Rsm  i}, \ \
Q_i^{-\Lsm}, \hspace{5.7cm}
\hbox{ dynamical generators}.
\eeq
}
Our aim in this paper is to find a field theoretical realization for generators in \rf{02092019-man-03},\rf {02092019-man-04}. We note that, with the exception of $J^{+-}$, the kinematical generators \rf{02092019-man-03} are quadratic in fields,%
\footnote{ The $J^{+-}$ takes the form $J^{+-} = G_0 + \irm x^+ P^-$, where the generator $G_0$ is quadratic in fields, while the light-cone Hamiltonian $P^-$ consists quadratic and higher order terms in fields.}
while, the dynamical generators \rf{02092019-man-04} consist quadratic and higher order terms in fields. To provide a field realization of generators of the extended Poincar\'e superalgebra, we use a light-cone gauge description of fields.

\noindent {\bf Content of component fields}. We now discuss component fields entering extended massless supermultiplets. To this end we use a label $\lambda$ to denote a helicity of a massless field, while the indices $i,j,k,l=1,\ldots,\NN$ stand for vector indices of the $su(\NN)$ algebra. Using such notation, we introduce a field $\phi_{\lambda\,;\,i_1\ldots i_q}$ which is (integer or half-integer) helicity-$\lambda$ field of the Poincar\'e algebra and rank-$q$ totally antisymmetric covariant tensor field of the $su(\NN)$ algebra.%
\footnote{ Transformations of the field $\phi_{\lambda;i_1\ldots i_q}$ under action of generators of the Poincar\'e algebra take the same form as the ones for the field $\phi_\lambda$ in (2.23)-(2.27) in Ref.\cite{Metsaev:2019dqt}.}
Now the field content entering arbitrary (integer or half-integer) spin $\NN$-extended massless supermultiplet of the Poincar\'e superalgebra in $R^{3,1}$ is given by
\beq
\label{02092019-man-05}
&& \{\lambda\}_\ext = \sum_{q=0,1,2,\ldots, \NN} \oplus \,\, \phi_{\lambda - \half q+\frac{1}{4}\NN\,;\, i_1\ldots i_q }(x)\,,
\\
&& \hspace{1.5cm} \lambda = \hbox{$0,\pm \half ,\pm 1,\pm \frac{3}{2},\ldots, \pm \infty$}\,, \hspace{1cm} \NN \in 4 \No\,,
\eeq
where
\beq
&& \phi_{\lambda;i_1\ldots i_q } \ \hbox{ are bosonic fields for } \ \lambda\in \Zo;
\nonumber\\[-14pt]
\label{02092019-man-05-a1} &&
\\[-14pt]
&& \phi_{\lambda;i_1\ldots i_q } \ \hbox{ are fermionic fields for } \lambda \in \Zo+\half\,.
\nonumber
\eeq
From \rf{02092019-man-05}, we see that the $\NN$-extended massless supermultiplet involves fields with the following values of the helicities
{\small
\beq
&& \lambda_\minrm,\,\,\lambda_\minrm+\hbox{$\half$}\,,\ldots\ldots, \lambda-\hbox{$\half$},\,\,\lambda,\,\,\lambda+\hbox{$\half$}, \ldots\ldots,\lambda_\maxrm - \hbox{$\half$}\,,\,\, \lambda_\maxrm\,;
\\
&& \lambda_\minrm = \lambda - \frac{1}{4}\NN\,, \qquad \lambda_\maxrm = \lambda + \frac{1}{4}\NN\,. \qquad
\eeq
}
Also, from \rf{02092019-man-05}, we see that multiplicity of the helicity $\lambda-\half q + \frac{1}{4}\NN$ is equal to $C_\NN^q$. We find it convenient to label the supermultiplet \rf{02092019-man-05} by the $\lambda$ instead of $\lambda_\maxrm$ (or $\lambda_\minrm$) because, by using such convention, a scalar supermultiplet is simply labelled  as $\{0\}_\ext$.

Fields \rf{02092019-man-05} depend on the space time-coordinates $x\equiv x^\pm,x^{\Rsm,\Lsm}$. By definition, fields \rf{02092019-man-05} satisfy the hermitian conjugation condition given by
\be \label{02092019-man-06}
\phi_{\lambda\,;\, i_1\ldots i_q }^\dagger(x) = \frac{(-)^{\frac{1}{4}\NN-\frac{1}{2}q  - \half e_{\frac{1}{2}q}}}{(\NN-q)!}\varepsilon^{i_1\ldots i_q i_{q+1}\ldots i_\NN} \phi_{-\lambda\,;\, i_{q+1}\ldots i_\NN }(x)\,,
\ee
where $\varepsilon^{i_1\ldots  i_\NN}$ stands for the Levy-Civita symbol of the $su(\NN)$ algebra, $\varepsilon^{1 \ldots \NN}=1$, while the quantity $e_\lambda$ is defined by the relations
\be \label{02092019-man-07}
e_\lambda =0 \hspace{0.5cm} \hbox{ for } \lambda \in \Zo\,,\hspace{1.4cm}
e_\lambda = 1 \hspace{0.5cm} \hbox{ for } \ \lambda \in \Zo+\half\,.
\ee

From now, in place of position-space fields \rf{02092019-man-05}, we prefer to use momentum-space fields which are defined by the Fourier transform with respect to the coordinates $x^-$, $x^\Rsm$, and $x^\Lsm$,
\be
\label{02092019-man-09}  \phi_{\lambda; i_1\ldots i_q}(x) = \int \frac{ d^3p }{ (2\pi)^{3/2} } e^{\irm(\beta x^- + p^\Rsm x^\Lsm  + p^\Lsm x^\Rsm)} \phi_{\lambda;i_1\ldots i_q}(x^+,p)\,,\qquad d^3p \equiv d\beta dp^\Rsm dp^\Lsm\,,
\ee
where we use the notation $\beta$ for the momentum in the plus light-cone direction $\beta\equiv p^+$. Note also that the argument $p$ of fields $\phi_{\lambda;i_1\ldots i_q}(x^+,p)$  stands as a shortcut for the momenta $\beta$, $p^\Rsm$, $p^\Lsm$. In terms of the momentum-space fields $\phi_{\lambda;i_1\ldots i_q}(x^+,p)$, the hermicity condition \rf{02092019-man-06} can be represented as
\be \label{02092019-man-10}
\phi_{\lambda\,;\, i_1\ldots i_q }^\dagger(p) = \frac{(-)^{\frac{1}{4}\NN-\half q  - \half e_{\half q}}}{(\NN-q)!}\varepsilon^{i_1\ldots i_q i_{q+1}\ldots i_\NN} \phi_{-\lambda\,;\, i_{q+1}\ldots i_\NN }(-p)\,.
\ee
Here and below, dependence of the momentum-space fields $\phi_{\lambda;i_1\ldots i_q}(p)$ on the evolution parameter $x^+$ is implicit. Let us also to note our convention $\phi_{\lambda;i_1\ldots i_q}^\dagger(p) \equiv( \phi_{\lambda;i_1\ldots i_q}(p))^\dagger$.

\noindent {\bf  Superfield formulation}.
To develop superfield formulation let us introduce a Grassmann-odd momentum $p_\theta^i$, $\{p_\theta^i,p_\theta^j\}=0$. The Grassmann momentum $p_\theta^i$ is a contravariant vector of the $su(\NN)$ algebra. The light-cone momentum superspace is parametrized by the light-cone evolution parameter $x^+$, the spatial momenta $p^\Rsm$, $p^\Lsm$, $\beta$ and the Grassmann momentum $p_\theta^i$,
\be \label{02092019-man-11}
x^+\,, \beta\,, \ p^\Rsm\,, \  p^\Lsm\,, \ p_\theta^i\,.
\ee
Using the Grassmann momentum $p_\theta^i$, we  collect component fields \rf{02092019-man-05},\rf{02092019-man-09} into superfield $\Phi_\lambda(p,p_\theta)$ defined as
\beq
\label{02092019-man-12} && \Phi_\lambda(p,p_\theta) = \sum_{q=0}^{\NN} \frac{1}{q!}\beta^{\frac{1}{4}\NN- \half q + \half e_\lambda - \half e_{\lambda - \half q } }\,\, p_\theta^{i_1}\ldots p_\theta^{i_q}  \phi_{\lambda - \half q+\frac{1}{4}\NN\,;\, i_1\ldots i_q }(p)\,,
\\
\label{02092019-man-12-a01} && \hspace{2cm} \lambda = \hbox{$0,\pm \half ,\pm 1,\pm \frac{3}{2},\ldots, \pm \infty$}\,, \hspace{1cm} \NN \in 4 \No\,,
\eeq
where $e_\lambda$ is defined in \rf{02092019-man-07}. Often, we use a shortcut $\Phi_\lambda\equiv \Phi_\lambda(p,p_\theta)$. From \rf{02092019-man-05-a1},\rf{02092019-man-12}, we see that
\beq
&& \Phi_\lambda \ \hbox{ are Grassmann even for } \ \lambda\in \Zo;
\nonumber\\[-16pt]
\label{02092019-man-12-a1} &&
\\[-16pt]
&& \Phi_\lambda  \ \hbox{ are Grassmann odd for } \ \lambda\in \hbox{$\Zo+\half$}\,.
\nonumber
\eeq

We note that, for  $\NN=4$, the scalar superfield $\Phi_0$ describes famous $\NN=4$ supersymmetric YM theory, while, for $\NN=8$, the scalar superfield $\Phi_0$ describes $\NN=8$ supergravity theory.

In order to obtain a field theoretical realization we need a realization of the $\NN$-extended Poincar\'e superalgebra in terms of differential operators. The realization  in terms of differential operators acting on our light-cone  superfield $\Phi_\lambda(p,p_\theta)$ takes the following form:
\beq
\label{02092019-man-14} && P^\Rsm = p^\Rsm\,,  \qquad P^\Lsm = p^\Lsm\,,   \hspace{2cm}   P^+=\beta\,,\qquad
P^- = p^-\,, \qquad p^-\equiv - \frac{p^\Rsm p^\Lsm}{\beta}\,,\qquad
\\
 \label{02092019-man-15}  && J^{+\Rsm}= \irm x^+ P^\Rsm + \partial_{p^\Lsm}\beta\,,
\hspace{2.4cm}  J^{+\Lsm}= \irm x^+ P^\Lsm + \partial_{p^\Rsm}\beta\,, \
\\
\label{02092019-man-16}  && J^{+-} = \irm x^+P^- + \partial_\beta \beta + M_\lambda^{+-}\,, \hspace{1cm} J^{\Rsm\Lsm} =  p^\Rsm\partial_{p^\Rsm} - p^\Lsm\partial_{p^\Lsm} + M_\lambda^{\Rsm\Lsm}\,,
\\
\label{02092019-man-17}  && J^{-\Rsm} = -\partial_\beta p^\Rsm + \partial_{p^\Lsm} p^-
+ M_\lambda^{\Rsm\Lsm}\frac{p^\Rsm}{\beta} - M_\lambda^{+-} \frac{p^\Rsm}{\beta}\,,
\\
\label{02092019-man-18}  && J^{-\Lsm} = -\partial_\beta p^\Lsm + \partial_{p^\Rsm} p^-
- M_\lambda^{\Rsm\Lsm}\frac{p^\Lsm}{\beta} - M_\lambda^{+-} \frac{p^\Lsm}{\beta}\,,
\\
\label{02092019-man-18-a1}  && \hspace{1.2cm} M_\lambda^{+-} =  \half p_\theta^i\partial_{p_\theta^i} - \frac{1}{4} \NN  - \half e_\lambda\,, \qquad M_\lambda^{\Rsm\Lsm}  = \lambda -\half p_\theta^i\partial_{p_\theta^i} +  \frac{1}{4} \NN\,,
\\
\label{02092019-man-19} && Q_i^{+\Rsm} = (-)^{e_\lambda} \beta \partial_{p_\theta^i}\,,
\hspace{2.3cm}  Q^{+\Lsm i} = (-)^{e_\lambda} p_\theta^i\,,
\\
\label{02092019-man-20} && Q^{-\Rsm i} =  (-)^{e_\lambda} \frac{1}{\beta} p^\Rsm p_\theta^i\,, \hspace{2cm} Q_i^{-\Lsm} = (-)^{e_\lambda} p^\Lsm \partial_{p_\theta^i}\,,
\\
\label{02092019-man-21} && J^i{}_j = p_\theta^i\partial_{p_\theta^j} - \frac{1}{\NN}\delta_j^i p_\theta^k \partial_{p_\theta^k}\,,
\\
&& \hspace{3cm}\partial_\beta \equiv \partial/\partial \beta\,, \hspace{0.5cm} \partial_{p^\Rsm} \equiv \partial/\partial p^\Rsm\,, \hspace{0.5cm}
\partial_{p^\Lsm} \equiv \partial/\partial p^\Lsm\,,
\eeq
where $\partial_{p_\theta^i}\equiv\partial/\partial p_\theta^i$ is left derivative with respect to the Grassmann momentum $p_\theta^i$.

To express hermicity condition in terms of the superfield we find it convenient to introduce new superfield $\Phi_\lambda^*$ defined by the relation
\be \label{02092019-man-22}
\Phi_\lambda^*(p,p_\theta) \equiv \beta^{\frac{\NN}{2}}\int d^\NN p_\theta^\dagger\, e^{ \frac{p_\theta^{i\dagger} p_\theta^i}{\beta} } (\Phi_\lambda(p,p_\theta))^\dagger\,.
\ee
It is easy to verify then that, in terms of the superfields $\Phi_\lambda$ and $\Phi_\lambda^*$, the hermicity condition \rf{02092019-man-10} takes the following simple form:
\be\label{02092019-man-23}
\Phi_{-\lambda}^*(-p,-p_\theta) =  \Phi_\lambda(p,p_\theta)\,.
\ee
Sometimes, we use a shortcut $\Phi_\lambda^*\equiv \Phi_\lambda^*(p,p_\theta)$. From \rf{02092019-man-23}, we see that the superfields $\Phi_\lambda$ and $\Phi_\lambda^*$ are not independent of each other. From \rf{02092019-man-12-a1} and \rf{02092019-man-23}, we note that the superfield
$\Phi_\lambda^*$ is Grassmann even for $\lambda\in \Zo$ and Grassmann odd for $\lambda\in \Zo+\half$.
Some helpful relations for the superfield $\Phi_\lambda^*$ may be found in Appendix B.

We now ready to provide a field theoretical realization of the Poincar\'e superalgebra.This is to say that, to quadratic order in fields, a field theoretical realization of the $\NN$-extended Poincar\'e superalgebra generators in terms of the superfields $\Phi_\lambda$ takes the following form:
\be \label{02092019-man-24}
G_\smpt  =  \sum_{\lambda=-\infty}^{+\infty} G_{\smpt,\,\lambda}  \qquad G_{\smpt,\, \lambda}  = \int d^3p\, d^\NN p_\theta \,\, \beta^{e_{\lambda+\half}} \Phi_\lambda^* G_{\diff,\,\lambda } \Phi_\lambda\,,
\ee
where a quantity $G_{\diff,\, \lambda}$ stands for the realization of the $\NN$-extended Poincar\'e superalgebra generators in terms of differential operators given in \rf{02092019-man-14}-\rf{02092019-man-21}.

By definition, the superfields $\Phi_\lambda$ and $\Phi_\lambda^*$ satisfy the Poisson-Dirac equal-time (anti)commutator given by
\be  \label{02092019-man-25}
[\Phi_\lambda(p,p_\theta),\Phi_{\lambda'}^*(p',p_\theta')]_\pm = \half \beta^{-e_{\lambda+\half}} \,\, \delta^3(p-p') \delta^\NN(p_\theta-p_\theta') \delta_{\lambda,\lambda'}\,,
\ee
where the notation $[a,b]_\pm$ is used for a graded commutator, $[a,b]_\pm=(-)^{\epsilon_a\epsilon_b+1}[b,a]_\pm$.
With the help of relations \rf{02092019-man-24},\rf{02092019-man-25}, we verify that the following equal-time (anti)commutator between the generators and the superfield $\Phi_\lambda$
\be  \label{02092019-man-26}
[\Phi_\lambda,G_\smpt]_\pm  =  G_{\diff,\,\lambda} \Phi_\lambda \,,
\ee
holds true, where the operators $G_{\diff,\,\lambda}$ are defined in \rf{02092019-man-14}-\rf{02092019-man-21}.

In conclusion of this section, we recall that the light-cone gauge action can be presented as
\be  \label{02092019-man-27}
S = \half\sum_{\lambda=-\infty}^\infty \int dx^+ d^3p d^\NN p_\theta \,\, \beta^{-e_\lambda} \Phi_\lambda^* \big( 2\irm \beta \partial^- - 2p^\Rsm p^\Lsm \big) \Phi_\lambda  +\int dx^+ P_{\rm int}^-\,,
\ee
where $\partial^-\equiv\partial/\partial x^+$, while $P_{\rm int}^-$ stands for light-cone gauge Hamiltonian that describes interacting fields.

\newsection{ \large $n$-point dynamical generators of $\NN$-extended Poincar\'e superalgebra} \label{sec-03}

As we have already noted we follow the method proposed in Ref.\cite{Dirac:1949cp} that reduces the problem of finding dynamical system to the problem of finding a solution of commutation relations for algebra of basic symmetries.
This implies that, for theories of interacting fields with extended supersymmetries in flat space, we should find interaction dependent deformation of the dynamical generators of the extended Poincar\'e superalgebra. In other words,  in theories of interacting fields, one has the following expansion in fields for the dynamical generators of the extended Poincar\'e superalgebra
\be \label{03092019-man-01}
G^\dyn
= \sum_{n=2}^\infty
G_\smpn^\dyn\,,
\ee
where $G_\smpn^\dyn$  \rf{03092019-man-01} is functional
that has $n$ powers of superfields $\Phi^*$.

Expressions for $G_\smpt^\dyn$ have been obtained in the previous section. Our aim in this Section is to discuss constraints on the dynamical generators $G_\smpn^\dyn$ with $n\geq 3$ which are obtained by using the kinematical symmetries of the Poincar\'e superalgebra. We describe the constraints in turn.

\noindent {\bf Kinematical $P^{\Rsm,\Lsm}$, $P^+$, $Q^{+\Lsm i}$ symmetries.}. Using (anti)commutators between the kinematical generators $P^\Rsm$, $P^\Lsm$, $P^+$, $Q^{+\Lsm i}$ and the dynamical generators \rf{02092019-man-04},  we find that the dynamical generators $G_\smpn^\dyn$ with $n\geq 3$ can be presented as:
\beq
\label{03092019-man-02} && P_\smpn^- = \int\!\! d\Gamma_\smpn\,\,  \langle \Phi_\smpn^*  | p_\smpn^-\rangle\,,
\\
\label{03092019-man-03} && Q_\smpn^{-\Rsm i} = \int\!\! d\Gamma_\smpn\,\,  \langle \Phi_\smpn^* | q_\smpn^{-\Rsm i}\rangle\,,
\\
\label{03092019-man-04} && Q_{i \smpn}^{-\Lsm} = \int\!\! d\Gamma_\smpn\,\,  \langle \Phi_\smpn^*  | q_{i \smpn}^{-\Lsm} \rangle\,,
\\
\label{03092019-man-05} && J_\smpn^{-\Rsm} = \int\!\! d\Gamma_\smpn\,\,  \langle\Phi_\smpn^* | j_\smpn^{-\Rsm}\rangle  +   \langle \Xbf_\smpn^\Rsm \Phi_\smpn^* | p_\smpn^-\rangle  - \langle \Xbf_{\theta\, i\,\smpn}  \Phi_\smpn^* |q_\smpn^{-\Rsm i} \rangle\,,
\\
\label{03092019-man-06} && J_\smpn^{-\Lsm} = \int\!\! d\Gamma_\smpn\,\,  \langle \Phi_\smpn^* | j_\smpn^{-\Lsm}\rangle +   \langle \Xbf_\smpn^\Lsm \Phi_\smpn^* | p_\smpn^- \rangle  + \frac{1}{n} \PP_{\theta\,\smpn}^i \langle \Phi_\smpn^*   | q_{i \smpn}^{-\Lsm} \rangle\,,
\eeq
where, in \rf{03092019-man-02}-\rf{03092019-man-06} and below, we use the following notation:
\beq
\label{03092019-man-07} && d\Gamma_\smpn = d\Gamma_\smpn^p d\Gamma_\smpn^{p_\theta} \,,
\\
\label{03092019-man-08} && d\Gamma_\smpn^p =  (2\pi)^3 \delta^{3}(\sum_{a=1}^n p_a)\prod_{a=1}^n \frac{d^3p_a}{(2\pi)^{3/2} }\,, \qquad d^3 p_a = dp_a^\Rsm dp_a^\Lsm d\beta_a\,,
\\
\label{03092019-man-09} && d\Gamma_\smpn^{p_\theta} \equiv  \delta^\NN(\sum_{a=1}^n  p_{\theta_a} ) \prod_{a=1}^n d^\NN p_{\theta_a}\,,
\\
\label{03092019-man-10} && \Xbf_\smpn^\Rsm =  - \frac{1}{n}\sum_{a=1}^n \partial_{p_a^\Lsm}\,, \hspace{1cm} \Xbf_\smpn^\Lsm = - \frac{1}{n}\sum_{a=1}^n\partial_{p_a^\Rsm}\,,
\\
\label{03092019-man-11} && \Xbf_{\theta i\,\smpn} = \frac{1}{n}\sum_{a=1}^n \partial_{p_{\theta_a}^i}\,,\hspace{1cm} \PP_{\theta\,\smpn}^i =  \sum_{a=1}^n \frac{p_{\theta_a}^i}{\beta_a}\,,
\eeq
and the index $a=1,\ldots,n$ is used to label superfields (and their arguments) entering $n$-point interaction vertex. We note also that, in \rf{03092019-man-02}-\rf{03092019-man-06}, we use the shortcuts  $\langle \Phi_\smpn^*| p_\smpn^-\rangle$, $\langle \Phi_\smpn^*| q_\smpn^{-\Rsm,\Lsm}\rangle$, and $\langle \Phi_\smpn^*| j_\smpn^{-\Rsm,\Lsm}\rangle$ for the following expressions
\beq
\label{03092019-man-12} && \langle \Phi_\smpn^*| p_\smpn^-\rangle \quad \equiv \sum_{\lambda_1\ldots\lambda_n} \Phi_{\lambda_1\ldots\lambda_n}^*  p_{\lambda_1\ldots\lambda_n}^-\,,
\\
\label{03092019-man-13} && \langle \Phi_\smpn^*| q_\smpn^{-\Rsm,\Lsm} \rangle \equiv \sum_{\lambda_1\ldots\lambda_n} \Phi_{\lambda_1\ldots\lambda_n}^*  q_{\lambda_1\ldots\lambda_n}^{-\Rsm,\Lsm}\,,
\\
\label{03092019-man-14} && \langle \Phi_\smpn^*| j_\smpn^{-\Rsm,\Lsm} \rangle \equiv \sum_{\lambda_1\ldots\lambda_n} \Phi_{\lambda_1\ldots\lambda_n}^*  j_{\lambda_1\ldots\lambda_n}^{-\Rsm,\Lsm}\,,
\\
\label{03092019-man-15} && \hspace{2cm} \Phi_{\lambda_1\ldots\lambda_n}^* \equiv  \Phi_{\lambda_1}^*(p_1,p_{\theta_1})  \ldots  \Phi_{\lambda_n}^*(p_n,p_{\theta_n}) \,.
\eeq
The quantities $p_{\lambda_1\ldots\lambda_n}^-$, $q_{\lambda_1\ldots\lambda_n}^{-\Rsm,\Lsm}$, and $j_{\lambda_1\ldots\lambda_n}^{-\Rsm,\Lsm}$ appearing in \rf{03092019-man-12}-\rf{03092019-man-14}, will be referred to as $n$-point densities. For brevity, we denote these densities as $g_{\lambda_1\ldots\lambda_n}$,
\be \label{03092019-man-16}
g_{\lambda_1\ldots\lambda_n} = p_{\lambda_1\ldots\lambda_n}^-,\quad
q_{\lambda_1\ldots\lambda_n}^{-\Rsm\,i},\quad
q_{i;\lambda_1\ldots\lambda_n}^{-\Lsm},\quad  j_{\lambda_1\ldots\lambda_n}^{-\Rsm},\quad
j_{\lambda_1\ldots\lambda_n}^{-\Lsm}\,.
\ee
In general, the densities $g_{\lambda_1\ldots\lambda_n}$ \rf{03092019-man-16} depend on the spatial momenta $p_a^\Rsm$, $p_a^\Lsm$, $\beta_a$, the Grassmann momenta $p_{\theta_a}^i$, and helicities $\lambda_a$, $a=1,2\ldots,n$,
\be \label{03092019-man-17}
g_{\lambda_1\ldots\lambda_n} = g_{\lambda_1\ldots\lambda_n}(p_a,p_{\theta_a})\,.
\ee
Note that the argument $p_a$ in delta-function \rf{03092019-man-08}, superfields \rf{03092019-man-15} and densities \rf{03092019-man-17} stands for the spatial momenta $p_a^\Rsm$, $p_a^\Lsm$, and $\beta_a$.
We note also that, in \rf{03092019-man-05},\rf{03092019-man-06}, the operators $\Xbf_\smpn^{\Rsm,\Lsm}$, $\Xbf_{\theta i\,\smpn}$ defined in \rf{03092019-man-10},\rf{03092019-man-11} are acting only on the arguments of the superfields. Namely, for example, the shortcut $\langle \Xbf_\smpn^\Rsm \Phi_\smpn^* | g_\smpn \rangle$ should read as follows
\be \label{03092019-man-18}
\langle \Xbf_\smpn^\Rsm \Phi_\smpn^* | g_\smpn\rangle =  \sum_{\lambda_1,\ldots \lambda_n} (\Xbf_\smpn^\Rsm\Phi_{\lambda_1\ldots\lambda_n}^* ) g_{\lambda_1\ldots\lambda_n}\,.
\ee

Often, we will refer to the density $p_\smpn^-$ as $n$-point interaction vertex, while, for $n=3$, the density $p_\smp3^-$ will be refereed to  as cubic interaction vertex.

\noindent {\bf $J^{+-}$-symmetry equations}. Commutators between the kinematical generator $J^{+-}$  and the dynamical generators $P^-$, $Q^{-\Rsm,\Lsm}$, $J^{-\Rsm,\Lsm}$ lead to the following equations for the densities:
\beq
\label{03092019-man-19} && \big( \Jbf^{+-} - \frac{(n-2)\NN}{4}\big) g_{\lambda_1\ldots \lambda_n} = 0\,, \hspace{1.6cm} \hbox{ for } \quad  g_{\lambda_1\ldots \lambda_n} =  p_{\lambda_1\ldots \lambda_n}^-\,,\,\, j_{\lambda_1\ldots \lambda_n}^{-\Rsm,\Lsm}\,,\qquad
\\
\label{03092019-man-20} && \big( \Jbf^{+-} - \frac{(n-2)\NN+2}{4}\big) g_{\lambda_1\ldots \lambda_n} = 0\,,  \hspace{0.8cm} \hbox{ for }  \quad g_{\lambda_1\ldots \lambda_n} =  q_{\lambda_1\ldots \lambda_n}^{-\Rsm i}\,,\,\, q_{i;\lambda_1\ldots \lambda_n}^{-\Lsm}\,,
\\
\label{03092019-man-21} && \hspace{3cm} \Jbf^{+-} \equiv \sum_{a=1}^n  \big( \beta_a\partial_{\beta_a} + \half p_{\theta_a}^i\partial_{p_{\theta_a}^i}  + \half e_{\lambda_a} \big)\,.
\eeq

\noindent {\bf $J^{\Rsm\Lsm}$-symmetry equations}. Commutators between the kinematical generator $J^{\Rsm\Lsm}$ and the dynamical generators $P^-$, $Q^{-\Rsm,\Lsm}$, $J^{-\Rsm,\Lsm}$ lead to the following equations for the densities:
\beq
\label{03092019-man-22} && \big(\Jbf^{\Rsm\Lsm} + \frac{(n-2)\NN}{4}\big) p_{\lambda_1\ldots \lambda_n}^- = 0 \,,
\\
\label{03092019-man-23} && \big(\Jbf^{\Rsm\Lsm} + \frac{(n-2)\NN-2}{4}\big) q_{\lambda_1\ldots \lambda_n}^{-\Rsm\,i} = 0 \,,
\\
\label{03092019-man-24} && \big(\Jbf^{\Rsm\Lsm} + \frac{(n-2)\NN+2}{4}\big) q_{i;\lambda_1\ldots \lambda_n}^{-\Lsm} = 0 \,,
\\
\label{03092019-man-25} && \big(\Jbf^{\Rsm\Lsm} + \frac{(n-2)\NN-4}{4}\big) j_{\lambda_1\ldots \lambda_n}^{-\Rsm} = 0 \,,
\\
\label{03092019-man-26} && \big(\Jbf^{\Rsm\Lsm} + \frac{(n-2)\NN+4}{4}\big) j_{\lambda_1\ldots \lambda_n}^{-\Lsm} = 0 \,,
\\
\label{03092019-man-27} && \hspace{3cm} \Jbf^{\Rsm\Lsm} \equiv \sum_{a=1}^n  \big( p_a^\Rsm\partial_{p_a^\Rsm} - p_a^\Lsm\partial_{p_a^\Lsm} - \half p_{\theta_a}^i\partial_{p_{\theta_a}^i} + \lambda_a \big)\,.
\eeq

\noindent {\bf $J^i{}_j$-symmetry equations}. Commutators between the kinematical generators $J^i{}_j$  and the dynamical generators $P^-$, $Q^{-\Rsm,\Lsm}$, $J^{-\Rsm,\Lsm}$ lead to the following equations for the densities:
\beq
\label{03092019-man-28} && \Jbf^i{}_j  g_{\lambda_1\ldots \lambda_n}    = 0\,, \hspace{1cm} \hbox{ for } \quad g_{\lambda_1\ldots \lambda_n} = p_{\lambda_1\ldots \lambda_n}^-\,, \,\, j_{\lambda_1\ldots \lambda_n}^{-\Rsm,\Lsm }\,,
\\
\label{03092019-man-29} && \Jbf^i{}_j  q_{\lambda_1\ldots \lambda_n}^{-\Rsm\, l} = \delta_j^l q_{\lambda_1\ldots \lambda_n}^{-\Rsm\, i} -\frac{1}{\NN}\delta_j^i q_{\lambda_1\ldots \lambda_n}^{-\Rsm\, l}\,,
\\
\label{03092019-man-30} && \Jbf^i{}_j  q_{l;\lambda_1\ldots \lambda_n}^{-\Lsm} = - \delta_l^i q_{j;\lambda_1\ldots \lambda_n}^{-\Lsm} + \frac{1}{\NN}\delta_j^i q_{l;\lambda_1\ldots \lambda_n}^{-\Lsm}\,,
\\
\label{03092019-man-31} && \hspace{3cm} \Jbf^i{}_j  \equiv \sum_{a=1}^n  \big( p_{\theta_a}^i\partial_{p_{\theta_a}^j} -\frac{1}{\NN} \delta_j^i p_{\theta_a}^k \partial_{p_{\theta_a}^k} \big)\,.
\eeq

\noindent {\bf $J^{+\Rsm}$, $J^{+\Lsm}$, $Q^{+\Rsm}$-symmetry equations}. Using (anti)commutators between kinematical generators $J^{+\Rsm}$, $J^{+\Lsm}$, and $Q^{+\Rsm}$ and the dynamical generators $P^-$, $Q^{-\Rsm,\Lsm}$, $J^{-\Rsm,\Lsm}$,  we verify that the densities $g_{\lambda_1\ldots\lambda_n}$ \rf{03092019-man-17}
can be presented as
\be \label{03092019-man-32}
g_{\lambda_1\ldots\lambda_n} = g_{\lambda_1\ldots\lambda_n} (\Po_{ab}^\Rsm,\Po_{ab}^\Lsm\,, \Po_{\theta\,ab},\beta_a)\,,
\ee
where we use the notation
\be \label{03092019-man-33}
\Po_{ab}^\Rsm \equiv p_a^\Rsm \beta_b - p_b^\Rsm \beta_a\,, \qquad
\Po_{ab}^\Lsm \equiv p_a^\Lsm \beta_b - p_b^\Lsm \beta_a\,, \qquad
\Po_{\theta\, ab}^i \equiv p_{\theta_a}^i \beta_b - p_{\theta_b}^i \beta_a\,.
\ee
In other words, the densities $g_{\lambda_1\ldots\lambda_n}$ \rf{03092019-man-17} turn out to be dependent on $\Po_{ab}^{\Rsm,\Lsm}$ and Grassmann momenta $\Po_{\theta\, ab}^i$ in place of the respective momenta $p_a^{\Rsm,\Lsm}$ and the Grassmann momenta $p_{\theta_a}^i$.

\noindent {\bf Restriction imposed by Grassmann parity}. In conclusion of this section, we note the following restriction on all densities in \rf{03092019-man-16}:
\be  \label{03092019-man-34}
g_{\lambda_1\ldots \lambda_n} = 0 \qquad \hbox{ for } \qquad \sum_{a=1}^n \lambda_a \in \hbox{$\Zo + \half$}\,.
\ee
Restriction \rf{03092019-man-34} is obtained by considering $J^i{}_j$ symmetries and Grassmann parity of the densities $g_{\lambda_1\ldots \lambda_n}$. Namely, on the one hand, in view of the $J^i{}_j$ symmetries, a dependence of the generators $P_\smpn^-$ and $J_\smpn^{-\Rsm,\Lsm}$ on the Grassmann momenta $\Po_{\theta\, ab}^i$ is realized by means of the Grassmann even quantities
\be
\varepsilon_{i_1 \ldots i_\NN}\Po_{a_1b_1}^{i_1} \ldots \Po_{a_\NN b_\NN}^{i_\NN}\,,
\ee
while, a dependence of the supercharges $Q_\smpn^{-\Rsm i}$ and $Q_{i\smpn}^{-\Lsm }$
on the Grassmann momenta $\Po_{\theta\, ab}^i$ is realized by means of the respective Grassmann odd quantities
\be
\Po_{\theta\, ab}^i\,, \qquad  \hspace{1cm} \varepsilon_{i i_2 \ldots i_\NN} \Po_{\theta\, a_2b_2}^{i_2} \ldots \Po_{\theta\, a_\NN b_\NN}^{i_\NN}\,.
\ee
On the other hand, by definition, the generators $P_\smpn^-$ and $J_\smpn^{-\Rsm,\Lsm}$ \rf{03092019-man-02}, \rf{03092019-man-05},\rf{03092019-man-06} should be Grassmann even, while the supercharges $Q_\smpn^{-\Rsm i}$ $Q_{i\smpn}^{-\Lsm }$ \rf{03092019-man-03},\rf{03092019-man-04} should be Grassmann odd. Taking into account above said and relations in \rf{02092019-man-12-a1}, we get the restriction \rf{03092019-man-34}.

We now proceed to the main theme of our study.

\newsection{ \large Complete system of equations for cubic vertices  } \label{sec-04}

In this Section, we present a complete system of equations required to determine the cubic interaction vertices  unambiguously. The complete system of equations is obtained by analysing the following three requirements.

\noindent 1) Kinematical symmetries.

\noindent 2) Dynamical symmetries.

\noindent 3) Light-cone gauge dynamical principle.

We now analyse these three requirements in turn.

\noindent {\bf Kinematical symmetries of cubic densities}. Kinematical symmetries for arbitrary $n$-point, $n\geq 3$, densities have already been considered in the previous section. For cubic densities, $n=3$, the kinematical symmetry equations can further be simplified in view of the following well known observation. Using the momentum conservation laws
\be  \label{04092019-man-01}
p_1^\Rsm + p_2^\Rsm + p_3^\Rsm = 0\,, \quad p_1^\Lsm + p_2^\Lsm + p_3^\Lsm = 0\,, \quad \beta_1 +\beta_2 +\beta_3 =0 \,,\quad p_{\theta_1}^i + p_{\theta_2}^i + p_{\theta_3}^i=0\,,
\ee
it is easy to see that six momenta $\Po_{12}^{\Rsm,\Lsm}$, $\Po_{23}^{\Rsm,\Lsm}$, $\Po_{31}^{\Rsm,\Lsm}$ and three Grassmann momenta $\Po_{\theta\, 12}^i$, $\Po_{\theta\, 23}^i$, $\Po_{\theta\, 31}^i$ \rf{03092019-man-33}  are expressed in terms of the respective two momenta $\Po^{\Rsm,\Lsm}$ and one Grassmann momentum $\Po_\theta^i$,
\be  \label{04092019-man-02}
\Po_{12}^{\Rsm,\Lsm} =\Po_{23}^{\Rsm,\Lsm} = \Po_{31}^{\Rsm,\Lsm} = \Po^{\Rsm,\Lsm} \,,\qquad
\Po_{\theta\, 12}^i =\Po_{\theta\, 23}^i = \Po_{\theta\, 31}^i = \Po_\theta^i \,,
\ee
where the new momenta $\Po^{\Rsm,\Lsm}$ and $\Po_\theta^i$ are defined as
\beq
&& \Po^\Rsm \equiv \frac{1}{3}\sum_{a=1,2,3} \betach_a p_a^\Rsm\,, \qquad \Po^\Lsm \equiv \frac{1}{3} \sum_{a=1,2,3} \betach_a p_a^\Lsm\,, \qquad
\nonumber\\
\label{04092019-man-04} && \Po_\theta^i \equiv \frac{1}{3}\sum_{a=1,2,3} \betach_a p_{\theta_a}^i\,, \qquad
\betach_a\equiv \beta_{a+1}-\beta_{a+2}\,, \quad \beta_a\equiv
\beta_{a+3}\,.
\eeq
Therefore, using the following simplified notation for the cubic densities:
\be  \label{04092019-man-05}
p_\smp3^- = p_{\lambda_1\lambda_2\lambda_3}^- \,, \qquad  q_\smp3^{-\Rsm\,i} = q_{\lambda_1\lambda_2\lambda_3}^{-\Rsm\,i}\,,\qquad  q_{i\smp3}^{-\Lsm} = q_{i;\lambda_1\lambda_2\lambda_3}^{-\Lsm}\,, \qquad  j_\smp3^{-\Rsm,\Lsm}  = j_{\lambda_1\lambda_2\lambda_3}^{-\Rsm,\Lsm}\,,
\ee
and taking into account relations \rf{03092019-man-32},\rf{04092019-man-02}, we see that the cubic densities $p_\smp3^-$, $q_\smp3^{-\Rsm,\Lsm}$, and $j_\smp3^{-\Rsm,\Lsm}$ depend on the momenta $\beta_a$, $\Po^{\Rsm,\Lsm}$, the Grassmann momentum $\Po_\theta^i$ and the helicities $\lambda_1$, $\lambda_2$, $\lambda_3$,
\beq
\label{04092019-man-06} && p_\smp3^- = p_{\lambda_1\lambda_2\lambda_3}^-(\Po^\Rsm,\Po^\Lsm,\Po_\theta, \beta_a)\,, \qquad  q_\smp3^{-\Rsm,\Lsm} = q_{\lambda_1\lambda_2\lambda_3}^{-\Rsm,\Lsm}(\Po^\Rsm,\Po^\Lsm,\Po_\theta, \beta_a)\,, \quad
\nonumber\\
&& j_\smp3^{-\Rsm,\Lsm} = j_{\lambda_1\lambda_2\lambda_3}^{-\Rsm,\Lsm}(\Po^\Rsm,\Po^\Lsm,\Po_\theta, \beta_a)\,.
\eeq
Now, restricting to the value $n=3$, we represent kinematical symmetry equations obtained in the previous section in terms of densities \rf{04092019-man-06}.

\noindent {\bf $J^{+-}$-symmetry equations}: Using \rf{04092019-man-06}, we find that, for $n=3$,  equations \rf{03092019-man-19}-\rf{03092019-man-21} can be represented as
\beq
\label{04092019-man-07} &&  \big(\Jbf^{+-} - \frac{1}{4}\NN \big) p_\smp3^-  = 0\,,
\\
\label{04092019-man-08} &&  \big(\Jbf^{+-} - \frac{1}{4}(\NN+2) \big) q_\smp3^{-\Rsm,\Lsm}  = 0\,,
\\
\label{04092019-man-09} &&  \big(\Jbf^{+-} - \frac{1}{4}\NN \big) j_\smp3^{-\Rsm,\Lsm}  = 0\,,
\eeq
where operator $\Jbf^{+-}$ is defined as
\beq
\label{04092019-man-10} && \Jbf^{+-} \equiv    N_{\Po^\Rsm} + N_{\Po^\Lsm}+ \frac{3}{2} N_{\Po_\theta} +  \sum_{a=1,2,3}   (\beta_a \partial_{\beta_a} + \half e_{\lambda_a})\,,
\\
\label{04092019-man-11} && N_{\Po^\Rsm} \equiv \Po^\Rsm\partial_{\Po^\Rsm}\,, \hspace{1cm} N_{\Po^\Lsm} \equiv \Po^\Lsm \partial_{\Po^\Lsm}\,, \hspace{1cm} N_{\Po_\theta} \equiv \Po_\theta^i \partial_{\Po_\theta^i}\,.\qquad
\eeq
\noindent {\bf $J^{\Rsm\Lsm}$-symmetry equations}: Using \rf{04092019-man-06}, we find that, for $n=3$,  equations \rf{03092019-man-22}-\rf{03092019-man-27} can be represented as
\beq
\label{04092019-man-12} &&  \big(\Jbf^{\Rsm\Lsm} + \frac{1}{4}\NN\big)  p_\smp3^-  = 0\,,
\\
\label{04092019-man-13} &&  \big(\Jbf^{\Rsm\Lsm} + \frac{1}{4}(\NN-2)\big)  q_\smp3^{-\Rsm\, i}  = 0\,,
\\
\label{04092019-man-14} &&  \big(\Jbf^{\Rsm\Lsm} + \frac{1}{4}(\NN+2)\big)  q_{i\,\smp3}^{-\Lsm}  = 0\,,
\\
\label{04092019-man-15} &&  \big(\Jbf^{\Rsm\Lsm} + \frac{1}{4}(\NN-4)\big)  j_\smp3^{-\Rsm}  = 0\,,
\\
\label{04092019-man-16} &&  \big(\Jbf^{\Rsm\Lsm} + \frac{1}{4}(\NN+4)\big)  j_\smp3^{-\Lsm}  = 0\,,
\eeq
where operator $\Jbf^{\Rsm\Lsm}$ is defined as
\be \label{04092019-man-17}
\Jbf^{\Rsm\Lsm} \equiv    N_{\Po^\Rsm} -  N_{\Po^\Lsm} - \half N_{\Po_\theta} +\Mbf_\lambda\,, \qquad \Mbf_\lambda \equiv \sum_{a=1,2,3} \lambda_a\,,
\ee
and we use the notation in \rf{04092019-man-11}.

\noindent {\bf $J^i{}_j$-symmetry equations}.  Using \rf{04092019-man-06}, we find that, for $n=3$,  equations \rf{03092019-man-28}-\rf{03092019-man-31} can be represented as
\beq
\label{04092019-man-18} && \Jbf^i{}_j  g_\smp3   = 0\,, \hspace{1cm} \hbox{ for } \quad g_\smp3 = p_\smp3^-\,, \,\, j_\smp3^{-\Rsm,\Lsm }\,,
\\
\label{04092019-man-19} && \Jbf^i{}_j  q_\smp3^{-\Rsm\, l} = \delta_j^l q_\smp3^{-\Rsm\, i} -\frac{1}{\NN}\delta_j^i q_\smp3^{-\Rsm\, l}\,,
\\
\label{04092019-man-20} && \Jbf^i{}_j  q_{l\,\smp3}^{-\Lsm} = - \delta_l^i q_{j\,\smp3}^{-\Lsm} + \frac{1}{\NN}\delta_j^i q_{l\,\smp3 }^{-\Lsm}\,,
\eeq
where operator $\Jbf^i{}_j$ is defined as
\be \label{04092019-man-21}  \Jbf^i{}_j  \equiv \Po_\theta^i \partial_{\Po_\theta^j} - \frac{1}{\NN} \delta_j^i \Po_\theta^k \partial_{\Po_\theta^k}\,.
\ee

We now proceed with studying the restrictions imposed by dynamical symmetries.

\noindent {\bf Dynamical symmetries of cubic densities}. Constraints on the cubic densities imposed by (anti) commutators  between the dynamical generators are referred to as dynamical symmetry constraints.  This is to say that the (anti)commutators to be considered are given by
\beq
\label{04092019-man-22}  && [P^-,J^{-\Rsm,\Lsm}]=0\,, \hspace{2.3cm}  [P^-,Q^{-\Rsm,\Lsm}]=0\,,
\\
\label{04092019-man-23} && [J^{-\Rsm},J^{-\Lsm}]=0\,, \hspace{2.4cm}  [Q^{-\Rsm,\Lsm},J^{-\Lsm}]=0\,,     \hspace{1.2cm}    [Q^{-\Rsm,\Lsm},J^{-\Rsm}]=0\,,
\\
\label{04092019-man-24} && \{Q^{-\Rsm i},Q_j^{-\Lsm} \} = - \delta_j^i P^-\,,\hspace{1cm}
\{Q^{-\Rsm i},Q^{-\Rsm j} \} = 0\,, \hspace{1cm} \{ Q_i^{-\Lsm},Q_j^{-\Lsm} \} = 0\,.\qquad
\eeq
First, we consider the commutators in \rf{04092019-man-22}. In the cubic approximation, the commutators \rf{04092019-man-22} take the form
\be \label{04092019-man-25}
[P_\smpt^- ,J_\smp3^{-\Rsm}] + [P_\smp3^-,J_\smpt^{-\Rsm}]=0\,, \qquad [P_\smpt^-,Q_\smp3^{-\Rsm,\Lsm}] + [P_\smp3^-, Q_\smpt^{-\Rsm,\Lsm}]=0\,.
\ee
We verify that equations \rf{04092019-man-25} allow us to express the densities $q_\smp3^{-\Rsm,\Lsm}$ and $j_\smp3^{-\Rsm,\Lsm}$ in terms of the cubic vertex $p_\smp3^-$ in the following way:
\beq
\label{04092019-man-26} && q_\smp3^{-\Rsm i} = - \frac{\Po_\theta^i}{\Po^\Lsm} p_\smp3^- \,,  \hspace{2.2cm}    q_{i \smp3}^{-\Lsm} =    \frac{\beta}{\Po^\Rsm} \partial_{\Po_\theta^i} p_\smp3^-  \,,
\\
\label{04092019-man-27} && j_\smp3^{-\Rsm}  = -\frac{\beta}{ \Po^\Rsm \Po^\Lsm } \Jbf^{-\Rsm} p_\smp3^- \,, \hspace{1cm} j_\smp3^{-\Lsm}  = -\frac{\beta}{ \Po^\Rsm \Po^\Lsm } \Jbf^{-\Lsm} p_\smp3^- \,,
\eeq
where operators $\Jbf^{-\Rsm}$, $\Jbf^{-\Lsm}$ are defined as
\beq
\label{04092019-man-28} && \Jbf^{-\Rsm} =   \frac{\Po^\Rsm}{\beta} \big( -\No_\beta + \Mo_\lambda  - \half \Eo_\lambda\big)\,,
\\
\label{04092019-man-29} && \Jbf^{-\Lsm}  =    \frac{\Po^\Lsm}{\beta} \big(-\No_\beta - \Mo_\lambda - \half \Eo_\lambda \big)\,,
\\
\label{04092019-man-30} && \hspace{1.3cm} \No_\beta = \frac{1}{3}\sum_{a=1,2,3}\betach_a \beta_a\partial_{\beta_a}\,, \hspace{1cm} \beta \equiv \beta_1\beta_2\beta_3\,,
\\
\label{04092019-man-31} &&  \hspace{1.3cm} \Mo_\lambda = \frac{1}{3}\sum_{a=1,2,3}\betach_a\lambda_a\,,\hspace{1cm} \Eo_\lambda = \frac{1}{3}\sum_{a=1,2,3}\betach_a e_{\lambda_a}\,,
\eeq
while the symbol $e_\lambda$ entering \rf{04092019-man-31} is given in \rf{02092019-man-07}.

Second, we verify that, if the dynamical symmetry equations for all densities \rf{04092019-man-26},\rf{04092019-man-27} and kinematical symmetry equations for the cubic vertex $p_\smp3^-$ \rf{04092019-man-07},\rf{04092019-man-12},\rf{04092019-man-18} are satisfied, then all kinematical symmetry equations for the densities $q_\smp3^{-\Rsm,\Lsm}$, $j_\smp3^{-\Rsm,\Lsm}$ are satisfied automatically.

Third, we verify that, if the dynamical symmetry equations for all densities \rf{04092019-man-26},\rf{04092019-man-27}
are satisfied, then all the dynamical symmetry equations obtained from (anti)commutators \rf{04092019-man-23},\rf{04092019-man-24} are satisfied automatically.

Thus, we see that the kinematical and dynamical symmetry constraints for cubic densities amount to equations for densities \rf{04092019-man-26},\rf{04092019-man-27} and equations for the cubic vertex $p_\smp3^-$ \rf{04092019-man-07},\rf{04092019-man-12},\rf{04092019-man-18}.
Equations \rf{04092019-man-26},\rf{04092019-man-27} and \rf{04092019-man-07},\rf{04092019-man-12},\rf{04092019-man-18} do not allow us to fix the cubic densities unambiguously. To determine the cubic densities unambiguously we need some additional requirement. We refer to such requirement as light-cone dynamical principle.

\noindent {\bf Light-cone gauge dynamical principle}. We formulate the light-cone gauge dynamical principle in the following way:

\noindent \ibf) Cubic densities $p_\smp3^-$, $q_\smp3^{-\Rsm,\Lsm}$, $j_\smp3^{-\Rsm,\Lsm}$ should be polynomial in the momenta  $\Po^\Rsm$, $\Po^\Lsm$;

\noindent \iibf) Cubic vertex $p_\smp3^-$ should respect the following constraint:
\be \label{04092019-man-32}
p_\smp3^-  \ne  \Po^\Rsm\Po^\Lsm W\,, \quad W \ \hbox{is polynomial in } \Po^\Rsm,\Po^\Lsm\,.
\ee
For the reader convenience, we note that the requirement in \ibf) is simply the light-cone counterpart of locality condition which is commonly used in Lorentz covariant formulations. We now comment on the constraint \rf{04092019-man-32}. As is well known, upon field redefinitions, the cubic vertex $p_\smp3^-$ for massless fields is changed by terms proportional to $\Po^\Rsm \Po^\Lsm$ (see, e.g., the discussion in Appendix B, in Ref.\cite{Metsaev:2005ar}). This implies that all cubic vertices that are proportional to $\Po^\Rsm \Po^\Lsm$ can be removed by using field redefinitions.  As cubic vertices $p_\smp3^-$ that can be removed by exploiting field redefinitions are out of our interest, we use the constraint \rf{04092019-man-32}.

\noindent {\bf Complete system of equations for cubic vertex}. We now present all equations we obtained for the cubic vertex. Namely, for the cubic vertex
\be  \label{04092019-man-33}
p_\smp3^- = p_{\lambda_1\lambda_2\lambda_3}^-(\Po^\Rsm,\Po^\Lsm,\Po_\theta, \beta_a)\,,
\ee
we found the following complete system of equations:
\beq
&& \hbox{\it Kinematical}  \quad  J^{+-},   \quad  J^{\Rsm\Lsm},\ \hbox{\it and}   \ J^i{}_j \ \hbox{\it symmetries};
\nonumber\\
\label{04092019-man-34}   && \big( \Jbf^{+-} - \frac{\NN}{4} \big) p_\smp3^- =0 \,, \hspace{2.5cm}  \big( \Jbf^{\Rsm\Lsm} +  \frac{\NN}{4} \big)p_\smp3^- = 0\,,
\hspace{1cm}   \Jbf^i{}_j p_\smp3^- = 0\,,
\\
&& \hbox{\it Dynamical} \quad P^-,\quad Q^{-\Rsm,\Lsm}, \ \hbox{\it and} \ J^{-\Rsm,\Lsm} \hbox{\it symmetries}
\nonumber\\
\label{04092019-man-35} &&  q_\smp3^{-\Rsm\,i} = - \frac{\Po_\theta^i}{\Po^\Lsm}  p_\smp3^-\,, \hspace{1.8cm}   q_{i\smp3}^{-\Lsm} = \frac{\beta}{\Po^\Rsm} \partial_{\Po_\theta^i} p_\smp3^-\,,  \hspace{1.8cm}   j_\smp3^{-\Rsm,\Lsm} = - \frac{\beta}{\Po^\Rsm\Po^\Lsm}\Jbf^{-\Rsm,\Lsm} p_\smp3^- \,,\hspace{2cm}
\\
&&\hbox{ \it Light-cone gauge dynamical principle:}
\nonumber\\
\label{04092019-man-36} && p_\smp3^-\,, \ q_\smp3^{-\Rsm,\Lsm}\,,  \ j_\smp3^{-\Rsm,\Lsm} \hspace{0.5cm} \hbox{ are polynomial in } \Po^\Rsm, \Po^\Lsm;
\\
\label{04092019-man-37} && p_\smp3^- \ne \Po^\Rsm\Po^\Lsm W, \hspace{1cm} W  \hbox{ is polynomial in } \Po^\Rsm, \Po^\Lsm; \qquad
\eeq
where operators $\Jbf^{+-}$, $\Jbf^{\Rsm\Lsm}$, $\Jbf^i{}_j$, and $\Jbf^{-\Rsm,\Lsm}$  are given in \rf{04092019-man-10},\rf{04092019-man-17},\rf{04092019-man-21} and \rf{04092019-man-28},\rf{04092019-man-29} respectively.

To conclude this Section, it is the equations given in \rf{04092019-man-34}-\rf{04092019-man-37} that constitute the complete system of equations which allow us to fix the cubic densities $p_\smp3^-$, $q_\smp3^{-\Rsm,\Lsm}$, $j_\smp3^{-\Rsm,\Lsm}$ unambiguously.
As a side of remark we note that by applying our complete system of equations to supersymmetric Yang-Mills and supergravity theories with extended supersymmetries, we verify that the complete system equations allows us to fix the cubic interactions of those supersymmetric theories unambiguously (up to coupling constants). We think therefore that it is worthwhile to apply our complete system of equations to study the cubic vertices of arbitrary spin $\NN$-extended supersymmetric theories.

\newsection{ \large Cubic interaction vertices } \label{sec-05}

Now we present the solution to our complete system of equations presented in \rf{04092019-man-34}-\rf{04092019-man-37}. Some details of solving these equations may be in Appendix C. This is to say that the general solution for the cubic vertex $p_{\lambda_1\lambda_2\lambda_3}^-$, the supercharges  $q_{\lambda_1\lambda_2\lambda_3}^{-\Rsm,\Lsm}$, and the angular momenta $j_{\lambda_1\lambda_2\lambda_3}^{-\Rsm,\Lsm}$  is given by
\beq
\label{05092019-man-01} && p_{\lambda_1\lambda_2\lambda_3}^- = V_{\lambda_1\lambda_2\lambda_3} + \Vb_{\lambda_1\lambda_2\lambda_3}\,,
\\
\label{05092019-man-02} && \hspace{1.5cm} V_{\lambda_1\lambda_2\lambda_3} =  C^{\lambda_1\lambda_2\lambda_3} (\Po^\Lsm)^{\frac{1}{4}\NN + \Mbf_\lambda } \prod_{a=1,2,3} \beta_a^{-\lambda_a - \half e_{\lambda_a}}\,,
\\
\label{05092019-man-03} && \hspace{1.5cm} \Vb_{\lambda_1\lambda_2\lambda_3}  = \Cb^{\lambda_1\lambda_2\lambda_3} (\Po^\Rsm)^{ \frac{1}{4}\NN - \Mbf_\lambda }\, (\varepsilon \Po_\theta^\NN)  \prod_{a=1,2,3} \beta_a^{ \lambda_a - \half \NN - \half e_{\lambda_a}}\,,
\\
\label{05092019-man-04} && q_{\lambda_1\lambda_2\lambda_3}^{-\Rsm\, i} = - C^{\lambda_1\lambda_2\lambda_3} (\Po^\Lsm)^{\frac{1}{4}\NN + \Mbf_\lambda -1}\, \Po_\theta^i \prod_{a=1,2,3} \beta_a^{-\lambda_a - \half e_{\lambda_a}  }\,,
\\
\label{05092019-man-05} && q_{i;\,\lambda_1\lambda_2\lambda_3}^{-\Lsm} =   \Cb^{\lambda_1\lambda_2\lambda_3} (\Po^\Rsm)^{\frac{1}{4}\NN - \Mbf_\lambda - 1} (\varepsilon \Po_\theta^{\NN-1})_i \prod_{a=1,2,3} \beta_a^{\lambda_a + 1 -  \half \NN - \half e_{\lambda_a}}\,,
\\
\label{05092019-man-06} && j_{\lambda_1\lambda_2\lambda_3}^{-\Rsm} =  - 2 C^{\lambda_1\lambda_2\lambda_3} \Mo_\lambda (\Po^\Lsm)^{\frac{1}{4}\NN + \Mbf_\lambda-1} \prod_{a=1,2,3} \beta_a^{-\lambda_a  - \half e_{\lambda_a}}\,,
\\
\label{05092019-man-07} && j_{\lambda_1\lambda_2\lambda_3}^{-\Lsm} =    2\Cb^{\lambda_1\lambda_2\lambda_3} \Mo_\lambda (\Po^\Rsm)^{\frac{1}{4}\NN - \Mbf_\lambda- 1 }\, (\varepsilon\Po_\theta^\NN) \prod_{a=1,2,3} \beta_a^{\lambda_a - \half \NN - \half e_{\lambda_a}}\,,\qquad
\eeq
where we use the notation
\beq \label{05092019-man-08}
&& \Mbf_\lambda = \sum_{a=1,2,3}\lambda_a\,, \hspace{1cm} \Mo_\lambda = \frac{1}{3}\sum_{a=1,2,3}\betach_a \lambda_a\,,
\\
\label{05092019-man-09} && (\varepsilon \Po_\theta^\NN)  \equiv \frac{1}{\NN!}  \varepsilon_{i_1\ldots i_\NN}  \Po_\theta^{i_1} \ldots \Po_\theta^{i_\NN}\,, \hspace{1cm}
(\varepsilon \Po_\theta^{\NN-1})_i \equiv \frac{1}{(\NN-1)!}  \varepsilon_{i i_2\ldots i_\NN}  \Po_\theta^{i_2} \ldots \Po_\theta^{i_\NN}\,.  \qquad
\eeq
Definition of the symbol $e_\lambda$ and momenta $\Po^{\Rsm,\Lsm}$, $\Po_\theta^i$, $\betach_a$ may be found in \rf{02092019-man-07} and \rf{04092019-man-04} respectively, while quantity $\varepsilon_{i_1\ldots \ldots i_\NN}$ is the Levy-Civita symbol of the $su(\NN)$ algebra, $\varepsilon_{1\ldots \ldots \NN}=1$. Quantities $C^{\lambda_1\lambda_2\lambda_3}$, $\Cb^{\lambda_1\lambda_2\lambda_3}$ entering our solution  \rf{05092019-man-01}-\rf{05092019-man-07} are coupling constants. In general, these coupling constants depend on the helicities $\lambda_1$, $\lambda_2$, $\lambda_3$. The coupling constants are nontrivial for the following values of $\NN$ and the superfield helicities $\lambda_1$, $\lambda_2$, $\lambda_3$:
\beq
\label{05092019-man-10}   && C^{\lambda_1\lambda_2\lambda_3} \ne 0\,, \hspace{ 1cm } \hbox{ for } \quad \frac{1}{4}\NN+ \Mbf_\lambda -1 \geq   0 \,, \hspace{1cm} \Mbf_\lambda \in \Zo\,;
\\
\label{05092019-man-11}  && \Cb^{\lambda_1\lambda_2\lambda_3 } \ne 0 \,, \hspace{ 1cm } \hbox{ for }\quad \frac{1}{4}\NN - \Mbf_\lambda - 1 \geq 0\,, \hspace{1cm} \Mbf_\lambda \in \Zo\,;
\\
\label{05092019-man-12}  && C^{\lambda_1\lambda_2\lambda_3 *} =  (-)^{\Mbf_\lambda} \Cb^{-\lambda_1-\lambda_2-\lambda_3} \,,
\eeq
where, in \rf{05092019-man-12}, the asterisk implies complex conjugation.
We make comments on the constraints for the coupling constants presented in  \rf{05092019-man-10} -\rf{05092019-man-12}.

\noindent \ibf) Constraint on $C^{\lambda_1\lambda_2\lambda_3}$ and first constraint on $\Mbf_\lambda$ and $\NN$ in \rf{05092019-man-10} are obtainable from the requirement the densities \rf{05092019-man-02},\rf{05092019-man-04}, and \rf{05092019-man-06} to be polynomial in  the momentum $\Po^\Lsm$. Accordingly, constraint on $\Cb^{\lambda_1\lambda_2\lambda_3}$ and first constraint on $\Mbf_\lambda$ and $\NN$ in \rf{05092019-man-11} are obtainable from the requirement the densities \rf{05092019-man-03},\rf{05092019-man-05}, and \rf{05092019-man-07} to be polynomial in  the momentum $\Po^\Rsm$.

\noindent \iibf) Constraint $\Mbf_\lambda\in \Zo$ in \rf{05092019-man-10},\rf{05092019-man-11}  is simply obtained from the one in \rf{03092019-man-34} when $n=3$.

\noindent \iiibf) Constraint on the coupling constants in \rf{05092019-man-12}  is obtained from the requirement the cubic Hamiltonian $P_\smp3^-$ to be hermitian. This constraint can straightforwardly  be derived by using relation \rf{07092019-man-13} in Appendix B.

To summarize,  relations \rf{05092019-man-10}-\rf{05092019-man-12} give the classification of cubic interaction vertices for $\NN$-extended massless arbitrary spin supermultiplets, while expressions \rf{05092019-man-01}-\rf{05092019-man-03} give the momentum superspace representation for these vertices.

\noindent {\bf Cubic interaction vertices in terms of component fields}. For the reader convenience, we now present cubic vertices in terms of the component fields.
To this end we focus on interaction of three superfields $\Phi_{\lambda_1}^*$, $\Phi_{\lambda_2}^*$, $\Phi_{\lambda_3}^*$ and represent the cubic Hamiltonian
in the following way:
\beq
\label{05092019-man-12-ad01} && P_\smp3^-(\Phi_{\lambda_1},\Phi_{\lambda_2}\Phi_{\lambda_3}) = \int d\Gamma_\smp3^p\,\, C^{\lambda_1\lambda_2\lambda_3} \Vbf^{ \Phi_{\lambda_1},\Phi_{\lambda_2}\Phi_{\lambda_3} } + h.c.\,,
\\
\label{05092019-man-12-ad02} && C^{\lambda_1\lambda_2\lambda_3} \Vbf^{\Phi_{\lambda_1}\Phi_{\lambda_2}\Phi_{\lambda_3} } \equiv  \int  d\Gamma_\smp3^{p_\theta}\,\, \Phi_{\lambda_1\lambda_2\lambda_3}^* V_{\lambda_1\lambda_2\lambda_3}\,,
\eeq
where expressions for $d\Gamma_\smp3^p$, $d\Gamma_\smp3^{p_\theta}$ are obtainable by setting $n=3$ in \rf{03092019-man-08},\rf{03092019-man-09}. It is the vertex $\Vbf^{\scriptscriptstyle\Phi_{\lambda_1}\Phi_{\lambda_2}\Phi_{\lambda_3} }$ \rf{05092019-man-12-ad02} that provides us the representation in terms of the component fields. To get explicit representation of $\Vbf^{\scriptscriptstyle\Phi_{\lambda_1}\Phi_{\lambda_2}\Phi_{\lambda_3} }$ in terms of component fields \rf{02092019-man-05} we plug \rf{05092019-man-02} into \rf{05092019-man-12-ad02} and use the representation for $\Phi_\lambda^*$ in terms of the component fields given in \rf{07092019-man-01-ad01}. Doing so, we get

\beq
\label{05092019-man-12-ad03} && \Vbf^{\Phi_{\lambda_1}\Phi_{\lambda_2}\Phi_{\lambda_3} } = \sum_{q_1,q_2,q_3=0\atop q_1+q_2+q_3=\NN}^\NN   C_{i(q_1)i(q_2)i(q_3)} V_{i(q_1)i(q_2)i(q_3)}^{\Lambda_1\Lambda_2\Lambda_3 }\,, \qquad
\\
\label{05092019-man-12-ad03-x1} && \hspace{3cm} V_{i(q_1)i(q_2)i(q_3)}^{\Lambda_1\Lambda_2\Lambda_3 } \equiv (\Po^\Lsm)^{\Lambda_1+ \Lambda_2+ \Lambda_3}  \prod_{a=1,2,3} \phi_{\Lambda_a; i(q_a)}^\dagger(p_a) \beta_a^{-\Lambda_a - \half e_{\Lambda_a}}\,, \qquad
\eeq
where we use the notation
\beq
\label{05092019-man-12-ad04} && \Lambda_a \equiv \lambda_a -\frac{q_a}{2} + \frac{\NN}{4}\,, \hspace{1cm} a=1,2,3\,,
\\
\label{05092019-man-12-ad05} && C_{ i(q_1)i(q_2)i(q_3)} \equiv  \frac{\omega_{q_1q_2q_3}}{q_1!q_2!q_3!}\int d\Gamma_\smp3^{p_\theta}
(\varepsilon p_{\theta_1}^{\NN-q_1})_{i(q_1)}  (\varepsilon p_{\theta_2}^{\NN-q_2})_{i(q_2)}
(\varepsilon p_{\theta_3}^{\NN-q_3})_{i(q_3)}\,,
\\
\label{05092019-man-12-ad06} && \hspace{2.2cm} \omega_{q_1q_2q_3} \equiv (-)^{e_{\lambda_1-\frac{q_1}{2} } e_{\frac{q_1}{2}} + \, e_{\lambda_3-\frac{q_3}{2} } e_{\frac{q_2}{2}}    } \,.
\eeq
In \rf{05092019-man-12-ad03}-\rf{05092019-man-12-ad05}, shortcut $i(q_a)$ stands for the $su(\NN)$ algebra tensor indices $i_1^a\ldots i_{q_a}^a$, while the quantities  $(\varepsilon p_\theta^{\NN-q})_{i(q)}$ appearing in \rf{05092019-man-12-ad05} are defined in \rf{06092019-man-18-ad01}. Also note that, in \rf{05092019-man-12-ad03},  the summation runs over those values of $q_1,q_2,q_3=0,1,\ldots,\NN$ which satisfy the restriction $q_1+q_2+q_3=\NN$. Such restriction is appearing in view of
\be \label{05092019-man-12-ad07}
C_{ i(q_1)i(q_2)i(q_3)} \ne 0 \ \ \ \hbox{ only for } \  \ \ \ q_1 + q_2+ q_3 = \NN\,,\qquad 0 \leq q_a \leq \NN\,, \quad a=1,2,3\,.
\ee

From \rf{05092019-man-12-ad03}, we see that our generic vertex $\Vbf^{\scriptscriptstyle\Phi_{\lambda_1}\Phi_{\lambda_2}\Phi_{\lambda_3} }$ is decomposed into elementary vertices denoted by $V_{\scriptscriptstyle i(q_1)i(q_2)i(q_3)}^{\scriptscriptstyle\Lambda_1\Lambda_2\Lambda_3 }$ \rf{05092019-man-12-ad03-x1}. We note that the elementary vertex $V_{\scriptscriptstyle i(q_1)i(q_2)i(q_3)}^{\scriptscriptstyle\Lambda_1\Lambda_2\Lambda_3 }$ describes interaction of three component fields  $\phi_{\scriptscriptstyle\Lambda_a; i(q_a)}^\dagger$, $a=1,2,3$, having the respective helicities $\Lambda_1$, $\Lambda_2$, and $\Lambda_3$

\noindent {\bf Internal symmetry}. Let us demonstrate the incorporation of internal symmetry in our model by considering the algebra $o(\Nsf)$ as internal symmetry algebra. The internal symmetry can then be incorporated into our model as follows.

First, in place of the superfields $\Phi_\lambda$, $\Phi_\lambda^*$, we use matrix-valued superfields $\Phi_\lambda^{\asf\bsf}$, $\Phi_\lambda^{*\asf\bsf}$, where indices $\asf,\bsf$ stand for matrix indices of the $o(\Nsf)$ algebra, $\asf,\bsf=1,\ldots,\Nsf$. By definition, our new matrix-valued superfields satisfy the following algebraic constraints
\be \label{05092019-man-13}
\Phi_\lambda^{\asf\bsf} = (-)^{\lambda + \frac{\NN}{4} +\half \eta_\lambda e_\lambda} \Phi_\lambda^{\bsf\asf} \,, \qquad \Phi_\lambda^{*\asf\bsf} = (-)^{\lambda + \frac{\NN}{4} + \half \eta_\lambda e_\lambda} \Phi_\lambda^{*\bsf\asf} \,, \qquad \eta_\lambda^2 = 1\,,\quad \eta_{-\lambda} =  - \eta_\lambda\,,
\ee
where $e_\lambda$ is given in \rf{02092019-man-07}. It is easy to check that the constraints \rf{05092019-man-13} are consistent in view of the relation $(-)^{2\lambda +\half  \NN + \eta_\lambda e_\lambda}=1$. Note that, in general, the sign of $\eta_\lambda$ may depend of $\NN$. As in the case of the singlet superfields \rf{02092019-man-23}, the superfields $\Phi_\lambda^{\asf\bsf}$ and $\Phi_\lambda^{*\asf\bsf}$ are related as
\be \label{05092019-man-14}
\Phi_{-\lambda}^{*\asf\bsf}(-p,-p_\theta) =  \Phi_\lambda^{\asf\bsf}(p,p_\theta)\,.
\ee

Second, in formulas for generators and the action \rf{02092019-man-24},\rf{02092019-man-27}, in place of $\Phi_\lambda^* \Phi_\lambda$, we  use $\Phi_\lambda^{*\asf\bsf} \Phi_\lambda^{\asf\bsf}$,
while, in the cubic vertices, in place of $\Phi_{\lambda_1}^*
\Phi_{\lambda_2}^* \Phi_{\lambda_3}^*$, we use the expressions
$\Phi_{\lambda_1}^{*\asf\bsf} \Phi_{\lambda_2}^{*\bsf\csf} \Phi_{\lambda_3}^{*\csf\asf}$.

Third, (anti)commutator \rf{02092019-man-25} is represented as
\beq
\label{05092019-man-15} &&  [\Phi_\lambda^{\asf\bsf}(p,p_\theta), \Phi_{\lambda'}^{*\asf'\bsf'}(p',p_\theta')]_\pm = \half \beta^{- e_{\lambda+\half}} \Pi_\lambda^{\asf\bsf,\asf'\bsf'} \delta^3(p-p')\delta^\NN(p_\theta-p_\theta')  \delta_{\lambda,\lambda'}\,,
\\
\label{05092019-man-16}  && \Pi_\lambda^{\asf\bsf,\asf'\bsf'} \equiv \half\big( \delta^{\asf\asf'} \delta^{\bsf\bsf'} + (-)^{\lambda + \frac{\NN}{4} + \half \eta_\lambda e_\lambda }  \delta^{\asf\bsf'} \delta^{\bsf\asf'} \big)\,, \qquad \Pi_\lambda^{\asf\bsf,\asf'\bsf'} \Pi_\lambda^{\asf'\bsf',\csf\esf} = \Pi_\lambda^{\asf\bsf,\csf\esf}\,.
\eeq

The following remarks are in order.

\noindent \ibf)  For $\lambda_1=0$, $\lambda_2=0$, $\lambda_3=0$, the vertex given in \rf{05092019-man-01} describes self-interacting scalar superfield $\Phi_0$ and such vertex has already been obtained in Ref.\cite{Bengtsson:1983pg}. Thus, our result for the cubic vertex
$p_{\lambda_1\lambda_2\lambda_3}^-$ given in \rf{05092019-man-01} agrees with previously reported result related to the particular values $\lambda_1=0$, $\lambda_2=0$, $\lambda_3=0$,
and provides expression for the cubic vertex $p_{\lambda_1\lambda_2\lambda_3}^-$ corresponding to arbitrary values of the superfield helicities $\lambda_1$, $\lambda_2$, $\lambda_3$.

\noindent \iibf) Our vertices $\Vbf^{\scriptscriptstyle \Phi_{\lambda_1}\Phi_{\lambda_2}\Phi_{\lambda_3} }$ \rf{05092019-man-12-ad03} can be considered as a supersymmetric completion of cubic vertices for bosonic massless fields in the $4d$ flat space found in Ref.\cite{Bengtsson:1986kh}.
We note however that a manifestly Lorentz covariant description of some light-cone gauge vertices presented in Ref.\cite{Bengtsson:1986kh} is not available so far. In Sec.6, in Ref.\cite{Metsaev:2019dqt}, we provided the detailed discussion of vertices in Ref.\cite{Bengtsson:1986kh} that can be translated into manifestly Lorentz covariant form. The reader interested in Lorentz covariant formulation of light-cone gauge vertices is invited to read Sec.6, in Ref.\cite{Metsaev:2019dqt}.

\noindent \iiibf) Taking into account relations \rf{05092019-man-12-ad04}, the restrictions on $q_a$ in \rf{05092019-man-12-ad07} can entirely be represented in terms of $\lambda_a$ and $\Lambda_a$ as,
\beq
\label{05092019-man-17}  && \lambda_1+\lambda_2 +\lambda_3 = \Lambda_1+\Lambda_2 +\Lambda_3 - \frac{1}{4}\NN\,,
\\
\label{05092019-man-17-ad01} && \Lambda_a - \frac{1}{4}\NN \leq \lambda_a \leq  \Lambda_a + \frac{1}{4}\NN \,, \qquad a =1,2,3\,.
\eeq

Restrictions \rf{05092019-man-17},\rf{05092019-man-17-ad01} provide the classification of  cubic interactions of the component fields which admit the supersymmetric completion. Namely, the cubic interactions of the three component fields having the helicities $\Lambda_1$, $\Lambda_2$, $\Lambda_3$ are described by the vertex in \rf{05092019-man-12-ad03-x1}. Restrictions \rf{05092019-man-17},\rf{05092019-man-17-ad01} tell us then which superfields $\Phi_\lambda$ are required for the supersymmetric completion of the vertex in \rf{05092019-man-12-ad03-x1}.
Also, from restrictions \rf{05092019-man-17},\rf{05092019-man-17-ad01}, we can learn which vertices in \rf{05092019-man-12-ad03-x1} do not admit supersymmetric completion.

For the reader convenience, we now illustrate the use of restrictions \rf{05092019-man-17},\rf{05092019-man-17-ad01}. To this end, for three spin-2 component fields, we consider cubic vertices of power $(\Po^\Lsm)^6$ in \rf{05092019-man-12-ad03-x1}. For spin-2 component fields, the helicities take values $\Lambda_a=\pm2$, $a=1,2,3$. From \rf{05092019-man-12-ad03-x1}, we see that, in order to get vertices of power $(\Po^\Lsm)^6$, we should choose $\Lambda_1=\Lambda_2=\Lambda_3=2$.  Plugging $\Lambda_a=2$, $a=1,2,3$, into \rf{05092019-man-17},\rf{05092019-man-17-ad01}, we obtain the restrictions
\beq
\label{05092019-man-17-ad02}  && \lambda_1+\lambda_2 +\lambda_3 = 6 - \frac{1}{4}\NN\,,
\\
\label{05092019-man-17-ad03} && 2 - \frac{1}{4}\NN \leq \lambda_a \leq  2 + \frac{1}{4}\NN \,, \qquad a =1,2,3\,.
\eeq
To explore further our illustrative example, we apply the restrictions \rf{05092019-man-17-ad02},\rf{05092019-man-17-ad03} to $\NN=8$ supergravity.
We recall that $\NN=8$ supergravity is described by the superfield $\Phi_\lambda$ with $\lambda=0$. Plugging $\NN=8$, $\lambda_1=\lambda_2=\lambda_3=0$ into \rf{05092019-man-17-ad02}, we see that the restriction \rf{05092019-man-17-ad02} is not satisfied. So, on the one hand, using \rf{05092019-man-17-ad02},\rf{05092019-man-17-ad03}, we are led to the well known statement: supersymmetries of $\NN=8$ supergravity do not admit supersymmetric completion of bosonic $R^3$-terms, where $R$ stands for the Riemann curvature tensor.
On the other hand, using \rf{05092019-man-17-ad02},\rf{05092019-man-17-ad03}, we can find superfields $\Phi_\lambda$ required for supersymmetric completion of the vertex of power $(\Po^\Lsm)^6$ for the three spin-2 fields. Obviously, to this end we should go beyond $\NN=8$ supergravity. Namely considering, for example, the particular case of the  superfields $\Phi_{\lambda_a}$, with $\lambda_1=6-\frac{1}{4}\NN$, $\lambda_2=\lambda_3=0$, and $\NN\geq 8$, we verify that restrictions \rf{05092019-man-17-ad02},\rf{05092019-man-17-ad03} are satisfied.

\noindent {\bf Conjecture for coupling constants of $\NN$-extended supersymmetric theory}. Let us set $\Phi_\lambda=0$ for $\lambda\in \Zo+\half$ in \rf{02092019-man-12},\rf{02092019-man-12-a01} and consider $\NN$-extended supersymmetric  model described by superfields $\Phi_\lambda$ with all $\lambda \in \Zo$. Using \rf{05092019-man-17}, we note that, if we choose the following solution for the cubic couplings constants:
\be \label{05092019-man-18}
C^{\lambda_1\lambda_2\lambda_3} = gk^{\lambda_1+\lambda_2 +\lambda_3 +\frac{1}{4}\NN} \big/(\lambda_1+\lambda_2 +\lambda_3 +\frac{1}{4}\NN - 1)!\,,
\ee
then, in terms of the helicities $\Lambda_a$ of the component fields appearing in \rf{05092019-man-12-ad03}, we get the relation
\be \label{05092019-man-19}
C^{\lambda_1\lambda_2\lambda_3} = gk^{\Lambda_1+\Lambda_2+\Lambda_3} \big/(\Lambda_1+\Lambda_2 +\Lambda_3 -1)!\,\,.
\ee
In \rf{05092019-man-18},\rf{05092019-man-19}, the $g$ is a dimensionless coupling constant, while the $k$ is a dimensionful parameter.
Relation for coupling constants \rf{05092019-man-19} coincides with one found in Refs.\cite{Metsaev:1991mt,Metsaev:1991nb} for bosonic theories of higher-spin fields. Thus, we see that for bosonic truncation of our $\NN$-extended supersymmetric model, solution given in \rf{05092019-man-18} coincides with the one in Refs.\cite{Metsaev:1991mt,Metsaev:1991nb}. Taking this into account, we then conjecture that generalization of our solution for coupling constants in Refs.\cite{Metsaev:1991mt,Metsaev:1991nb} to the case of $\NN$-extended supersymmetric model is given by the relation in \rf{05092019-man-18}.%
\footnote{ Solution for the cubic coupling constants \rf{05092019-man-19} of bosonic higher-spin theories was found in Refs.\cite{Metsaev:1991mt,Metsaev:1991nb} by analyzing the quartic approximation. In order to prove our conjecture for the cubic coupling constants \rf{05092019-man-18} one needs to extend analysis of cubic approximation in this paper to the  quartic approximation for the $\NN$-extended supersymmetric higher-spin theories. As a side remark we note that, taking into account relation \rf{05092019-man-17}, it is easy to see that the solution \rf{05092019-man-18} is unique solution that leads to \rf{05092019-man-19}.}
Also one can conjecture that solution \rf{05092019-man-18} supplemented by the constraint $\Cb^{\lambda_1\lambda_2\lambda_3}=0$ provides $\NN$-extended supersymmetric generalization of the bosonic higher-spin chiral model in Ref.\cite{Ponomarev:2016lrm}.

\newsection{ \large Conclusions}\label{concl}

In this paper, we generalized our previous study of $\NN=1$ massless arbitrary spin supermultiplets in the flat $4d$ space in Ref.\cite{Metsaev:2019dqt} to the case of $\NN$-extended massless arbitrary spin supermultiplets, $\NN=4\No$. For the $\NN$-extended massless supermultiplets, we built unconstrained superfields and used such superfields to develop the light-cone gauge superspace formulation.
We used our light-cone gauge superfield formulation to get  full list of the cubic interaction vertices for $\NN$-extended massless arbitrary spin (integer and half-integer) supermultiplets. We obtained restrictions on the values of $\NN$ and helicities of superfields which provide the complete classification of cubic vertices for the $\NN$-extended massless supermultiplets studied in this paper. We note also that our treatment of light-cone gauge superfields provides us the attractively simple superspace representation for the cubic interaction vertices. Now we would like to discuss potentially interesting generalizations and applications of our study.

\medskip

\noindent \ibf) Perhaps most interesting generalization of our results in this paper is related to the light-cone gauge higher-spin field theory in AdS. Light-cone gauge formulation of interacting higher-spin massless fields in $AdS_4$ space has recently been developed in Ref.\cite{Metsaev:2018xip}. Namely, in Ref.\cite{Metsaev:2018xip}, we demonstrated that the flat space cubic bosonic vertices obtained in Ref.\cite{Bengtsson:1986kh} enter as building blocks into AdS cubic bosonic vertices.
We expect therefore that results, methods, and approaches in this paper and in Ref.\cite{Metsaev:2018xip} will have interesting applications for studying light-cone gauge $\NN$-extended supersymmetric theories in $AdS_4$ space. For example, in Ref.\cite{Metsaev:2018xip}, we shown that the flat light-cone gauge bosonic vertices are in one-to-one correspondence with the AdS light-cone gauge bosonic vertices. For supersymmetric light-cone gauge flat and AdS cubic vertices, we also expect the one-to-one correspondence. This implies then that our classification for the $\NN$-extended flat cubic vertices obtained in this paper provides immediately the classification for $\NN$-extended AdS cubic vertices.%
\footnote{ We think that results in this paper might also have interesting applications for the studying supersymmetric extension of the conjectured non-local higher-spin field theories in flat space discussed in Ref.\cite{Roiban:2017iqg}.}
Here, for the reader convenience, we note that Vasiliev's equations for higher-spin gauge fields in $AdS_4$ were obtained in Ref.\cite{Vasiliev:1990en}. The complete
cubic coupling was found in Ref.\cite{Sleight:2016dba} and the quartic interaction was reconstructed in Refs.\cite{Bekaert:2015tva,Sleight:2017fpc}. Recent development of approach in Ref.\cite{Vasiliev:1990en} may be found in Refs.\cite{Didenko:2018fgx}.
In the framework of approach in Ref.\cite{Vasiliev:1990en}, various $\NN$-extended supersymmetric higher-spin gauge field theories in AdS space were studied in Refs.\cite{Konstein:1989ij}-\cite{Alkalaev:2002rq}. Also we mention the metric-like approaches in Refs.\cite{Joung:2012fv}-\cite{Karapetyan:2019psg} which might be helpful for studying higher-spin supersymmetric theories in AdS. Application of collective dipole approach for
the investigation of higher-spin interactions may be found in Refs.\cite{Koch:2010cy}.
We expect that light-cone gauge approach will be helpful for investigation of the problem of bulk definition of higher-spin theories identified in Ref.\cite{Sleight:2017pcz}.

\noindent \iibf) In this paper, we restricted our study to massless  supermultiplets in the four dimensions. Generalization of our study to the case of massless supermultiplets in the higher dimensions could be of interest.
In this respect, we note that all parity invariant cubic vertices for massless arbitrary spin light-cone gauge bosonic and fermionic fields in the higher dimensions  were built in Refs.\cite{Metsaev:2005ar,Metsaev:2007rn}, while the discussion of massless supermultiplets in higher dimensions may be found, e.g., in Ref.\cite{Sorokin:2018djm}.%
\footnote{ In the framework of BRST-BV approach and various metric-like Lorentz covariant approaches, cubic interactions for massless higher-spin fields were investigated in the respective Refs.\cite{Bekaert:2005jf}-\cite{Henneaux:2012wg} and Refs.\cite{Manvelyan:2010jr}.
Lorentz covariant parity-odd cubic interactions for higher-spin massless fields in $R^{3,1}$ are studied in Ref.\cite{Conde:2016izb}.
Recent interesting studies of fermionic fields may be found in Ref.\cite{Khabarov:2019dvi}.}
We expect therefore that studies in Refs.\cite{Metsaev:2005ar,Metsaev:2007rn,Sorokin:2018djm} might be helpful for the investigation of interacting supermultiplets in the higher dimensions.%
\footnote{ Twistor methods addressed, e.g., in Refs.\cite{Bandos:2019zqp,Uvarov:2018ose}, could also be helpful for studying interactions of massless supermultiplets  in higher dimensions.}

\noindent \iiibf) We expect that our results for supersymmetric {\it massless} higher-spin light-cone gauge fields obtained in this paper might be helpful for the extension of our study to the case of supersymmetric {\it massive} fields.
In light-cone gauge approach, interaction vertices for massive arbitrary spin bosonic and fermionic fields in the flat space were studied in Refs.\cite{Metsaev:2005ar,Metsaev:2007rn}. We think that light-cone gauge cubic vertices in Refs.\cite{Metsaev:2005ar,Metsaev:2007rn} will be helpful for the studying supersymmetric theories of massless and massive fields. For the reader convenience, we note that, in Lorentz covariant approach, $\NN=1$ higher-spin massless supermultiplets, by using BRST method, were studied in Ref.\cite{Buchbinder:2015kca}, while the $\NN=1$
massive supermultiplets are considered in Ref.\cite{Zinoviev:2007js}. Cubic self-interactions of massive fields and couplings of massive fields to massless fields were studied by using BRST approach in Ref.\cite{Metsaev:2012uy}.

\noindent \ivbf) In the recent time, higher-spin theories in three-dimensional flat and AdS spaces have extensively been studied in the literature. Namely, we mention that the interacting massless higher-spin gauge fields in $3d$ flat space have been studied in Ref.\cite{Mkrtchyan:2017ixk}, while massive higher-spin fields have been investigated in Refs.\cite{Buchbinder:2017izy}-\cite{Kuzenko:2016qwo}.
Recent applications of conformal geometry for studying $3d$ conformal higher-spin fields may be found in Refs.\cite{Henneaux:2018agj,Kuzenko:2019ill}, while unfolded formulation of $3d$ conformal fields is considered in Refs.\cite{Nilsson:2015pua}. We note, because the massless light-cone gauge higher-spin fields are trivial in $3d$ space, the usefulness of the light-cone formalism for studying such fields is not obvious. However, for the case of massive fields and conformal fields, we expect that the light-cone gauge approach might be helpful for better understanding of various aspect of massive and conformal field theories in three dimensions.
For the reader convenience, we note that light-cone formulation of higher-spin  massive fields in the $3d$ flat space is well known, while the light-cone gauge formulation of higher-spin massive fields in $AdS_3$ was obtained in Refs.\cite{Metsaev:1999ui,Metsaev:2000qb}. In the framework of ordinary-derivative (2nd-derivative) light-cone gauge formalism, higher-spin conformal fields were studied in Ref.\cite{Metsaev:2016rpa}.

\noindent \vbf) Quantum properties of bosonic higher-spin gauge field theories were studied in Refs.\cite{Ponomarev:2016jqk,Skvortsov:2018jea}. In Ref.\cite{Skvortsov:2018jea}, the arguments were given for UV finiteness of bosonic chiral higher-spin theory. We note also that, in the framework of light-cone approach, recent discussion of quantum properties of $\NN=8$ supergravity may be found in Ref.\cite{Kallosh:2009db}. We believe that our results for cubic interactions of $\NN$-extended arbitrary spin supermultiplets and methods in Refs.\cite{Kallosh:2009db,Skvortsov:2018jea} might be helpful for study of quantum properties of $\NN$-extended supersymmetric higher-spin field theories.
As note in the literature, extended $\NN=8$ supergravity theory is a candidate for UV finite theory (see, e.g., Ref.\cite{Kallosh:2009db} and references therein). We think therefore that supersymmetric (chiral and non-chiral) higher-spin theories are also candidates for UV finite theories. Last but not least motivation for our interest in supersymmetric higher-spin theories is related to the fact that supersymmetry makes study of four point vertices easier. We expect that, as compared to bosonic higher-spin theories, interesting features of the supersymmetric higher-spin theories will be seen upon consideration of four point vertices. For the case of 11d supergravity, example of application of supersymmetry for the studying four point vertices can be found in Sec.5 in Ref.\cite{Metsaev:2004wv}.

\noindent \vibf) Application of light-cone gauge approach for studying interacting continuous-spin bosonic field may be found in Ref.\cite{Metsaev:2017cuz,Metsaev:2018moa}. We expect that the methods developed in this paper might be helpful for studying interactions of supersymmetric continuous-spin fields. In the Lorentz covariant frame, the study of interactions of bosonic continuous-spin field may be found in Refs.\cite{Bekaert:2017xin}. Discussion of light-cone gauge continuous-spin field in AdS is given in Refs.\cite{Metsaev:2017myp,Metsaev:2019opn}.

\medskip

{\bf Acknowledgments}. This work was supported by the RFBR Grant No.17-02-00546.

\setcounter{section}{0}\setcounter{subsection}{0}
\appendix{ \large Notation and conventions  }

{\bf $\NN$-extended Poincar\'e superalgebra in light-cone basis}. Using notation $P^\mu$, $J^{\mu\nu}$, $\mu,\nu=0,1,2,3$, for generators of the Poincar\'e algebra, we present (anti)commutators of the $\NN$-extended Poincar\'e superalgebra given in \rf{02092019-man-03},\rf{02092019-man-04} as

\noindent { \it Commutators of generators of Poincar\'e algebra,  $P^\mu$, $J^{\mu\nu}$, and generators of $su(\NN)$ algebra, $J^i{}_j$}:
\beq
\label{06092019-man-01} && [P^\mu,\,J^{\nu\rho}]=\eta^{\mu\nu} P^\rho - \eta^{\mu\rho} P^\nu\,,
\qquad {} [J^{\mu\nu},\,J^{\rho\sigma}] = \eta^{\nu\rho} J^{\mu\sigma} + 3\hbox{ terms}\,,
\\
\label{06092019-man-02} && [J^i{}_j,J^k{}_l] = \delta_j^k J^i{}_l - \delta_l^i J^k{}_j\,, \hspace{1.5cm} i,j,k,l = 1,\ldots, \NN\,;
\eeq

\noindent { \it Commutators between supercharges}:
\beq
\label{06092019-man-03} && \{ Q_j^{+\Rsm},Q^{+\Lsm i} \} = \delta_j^i P^+\,, \qquad \{ Q^{-\Rsm i},Q_j^{-\Lsm} \} = - \delta_j^i  P^-\,,
\\
\label{06092019-man-04} && \{ Q_j^{+\Rsm},Q^{-\Rsm i} \} = \delta_j^i P^\Rsm\,, \qquad   \{Q^{+\Lsm i},Q_j^{-\Lsm}\} = \delta_j^i P^\Lsm\,;
\eeq

\noindent { \it Commutators between supercharges and generators of Lorentz algebra}:
\beq
\label{06092019-man-05} && [J^{+-},Q^{\pm \Rsm}] = \pm \half Q^{\pm \Rsm}\,,\qquad [J^{+-},Q^{\pm \Lsm}] = \pm \half Q^{\pm \Lsm}\,,
\\
\label{06092019-man-06} && [J^{\Rsm\Lsm},Q^{\pm \Rsm}] = \half Q^{\pm \Rsm}\,,\qquad \quad [J^{\Rsm\Lsm},Q^{\pm \Lsm}] = - \half Q^{\pm \Lsm}\,,
\\
\label{06092019-man-07} && [Q^{-\Rsm i},J^{+\Lsm}] = - Q^{+\Lsm i}\,, \hspace{1cm} [Q_i^{-\Lsm},J^{+\Rsm}] = - Q_i^{+\Rsm}\,,
\\
\label{06092019-man-08} && [Q_i^{+\Rsm},J^{-\Lsm}] =   Q_i^{-\Lsm}\,, \hspace{1.5cm} [Q^{+\Lsm i},J^{-\Rsm}] =  Q^{-\Rsm i}\,;
\eeq

\noindent {\it Commutators between supercharges and generators of $su(\NN)$ algebra}:
\beq
\label{06092019-man-09} [Q_i,J^j{}_k] = \delta_i^j Q_k - \frac{1}{\NN}\delta_k^j Q_i\,, \qquad [Q^i,J^j{}_k] = -\delta_k^i Q^j  + \frac{1}{\NN} \delta_k^j Q^i\,.
\eeq

In the light-cone basis \rf{02092019-man-03},\rf{02092019-man-04}, commutation relations for generators  of the Poincar\'e algebra can be obtained from \rf{06092019-man-01} by using the flat metric $\eta^{\mu\nu}$ which has the following non-vanishing elements $\eta^{+-}=\eta^{-+}=1$, $\eta^{\Rsm\Lsm}=\eta^{\Lsm\Rsm}=1$. Also note that, in \rf{06092019-man-09}, the shortcut $Q_i$ is used to indicate the supercharges $Q_i^{+\Rsm}$, $Q_i^{-\Lsm}$, while the shortcut $Q^i$ is used to indicate the supercharges $Q^{+\Lsm i}$, $Q^{-\Rsm i}$.

Hermitian properties of the generators are assumed to be as follows:
\beq
\label{06092019-man-10} && \hspace{-1.4cm}
P^{\pm \dagger} = P^\pm, \qquad \ \
P^{\Rsm\dagger} = P^\Lsm, \qquad
J^{\Rsm\Lsm\dagger} =  J^{\Rsm\Lsm}\,,\quad
J^{+-\dagger} = - J^{+-}, \quad
J^{\pm \Rsm\dagger} = -J^{\pm \Lsm}\,,
\nonumber\\
&& \hspace{-1.4cm}
Q_i^{+\Rsm\dagger} = Q^{+\Lsm i}\,, \hspace{0.6cm}  Q^{-\Rsm i\dagger} = Q_i^{-\Lsm}\,, \qquad J^i{}_j{}^\dagger = J^j{}_i\,.
\eeq
Covariant and contravariant vectors fields $X_i$, $X^i$ of the $su(\NN)$ algebra are transformed as
\be \label{06092019-man-11}
[X_i,J^j{}_k] = \delta_i^j X_k - \frac{1}{\NN}\delta_k^j X_i\,, \qquad [X^i,J^j{}_k] = -\delta_k^i X^j  + \frac{1}{\NN} \delta_k^j X^i\,.
\ee
Hermitian conjugated of the field $\phi_{\lambda;i_1\ldots i_q}(p)$ \rf{02092019-man-09} is denoted as $\phi_{\lambda;i_1\ldots i_q}^\dagger(p)$. Note that the fields $\phi_{\lambda;i_1\ldots i_q}(p)$ and $\phi_{\lambda;i_1\ldots  i_q}^\dagger(p)$ are the respective covariant and contravariant tensor fields of the $su(\NN)$ algebra. Transformations of $\phi_{\lambda;i_1\ldots i_q}(p)$ and $\phi_{\lambda;i_1\ldots  i_q}^\dagger(p)$  under action of the generators of the $su(\NN)$ algebra are realized as tensor products of the respective transformations for covariant and contravariant vector fields $X_i$ and $X^i$ \rf{06092019-man-11}.

\noindent {\bf Grassmann algebra}. Grassmann momentum is denoted by $p_\theta^i$. Throughout this paper we use the left derivative $\partial_{p_\theta^i}$ w.r.t the Grassmann momentum $p_\theta^i$, $\partial_{p_\theta^i} p_\theta^j=\delta_i^j$.
The integral over the Grassmann momentum $p_\theta^i$ is normalized to be
\be \label{06092019-man-12}
\int d^\NN p_\theta\,  p_\theta^{i_1} \ldots p_\theta^{i_\NN} =\varepsilon^{i_1\ldots i_\NN}\,,
\ee
where $\varepsilon^{i_1\ldots \ldots i_\NN}$ is the Levy-Civita symbol of the $su(\NN)$ algebra, $\varepsilon^{1\ldots \ldots \NN}=1$.
Ghost parities of the $p_\theta^i$, $\partial_{p_\theta^i}$, and measure $d^\NN p_\theta$ are given by
\be \label{06092019-man-13}
\GP({p_\theta^i) = 1\,, \qquad \GP(\partial_{p_\theta^i}) = 1\,, \qquad
\GP(d^\NN p_\theta}) = 0\,.
\ee
We note the following relations which are helpful for analysis of supercharges
\beq
\label{06092019-man-14} && p_\theta^i (\varepsilon p_\theta^{\NN-1})_j = \delta_j^i (\varepsilon p_\theta^\NN)\,, \hspace{2.2cm} \partial_{p_\theta^i} (\varepsilon p_\theta^\NN) = (\varepsilon p_\theta^{\NN-1})_i\,,
\\
\label{06092019-man-15} && (\varepsilon p_\theta^\NN)  \equiv \frac{1}{\NN!}  \varepsilon_{i_1\ldots i_\NN}  p_\theta^{i_1} \ldots p_\theta^{i_\NN}\,, \hspace{1cm}
(\varepsilon p_\theta^{\NN-1})_i \equiv \frac{1}{(\NN-1)!}  \varepsilon_{i i_2\ldots i_\NN}  p_\theta^{i_2} \ldots p_\theta^{i_\NN}\,.  \qquad
\eeq

The hermitian conjugation for product of two quantities $A$, $B$ having arbitrary ghost numbers is defined according to the rule $(AB)^\dagger = B^\dagger A^\dagger$.  For the Berezin integral, we use  the rule
\be \label{06092019-man-16}
\int d^\NN p_\theta\, (\partial_{p_\theta^i} A) B =  (-)^{\epsilon_A+1} \int d^\NN p_\theta  A  \partial_{p_\theta^i} B\,,  \qquad \epsilon_A \equiv \GP(A)\,.
\ee

Grassmann Dirac delta-function is defined by the relations
\be \label{06092019-man-17}
\delta^\NN(p_\theta) = p_\theta^1\ldots p_\theta^\NN\,, \qquad \int d^\NN p_\theta'\, \delta^\NN(p_\theta'-p_\theta) f(p_\theta') = f(p_\theta)\,,
\ee
while the Grassmann Fourier transformation and its inverse are fixed to be
\be \label{06092019-man-18}
F(p_\theta) = \int d^\NN p_\theta'e^{\frac{p_\theta' p_\theta}{\beta} } f(p_\theta')\,, \qquad f(p_\theta) = \beta^\NN \int d^\NN p_\theta'e^{\frac{p_\theta' p_\theta}{\beta} } F(p_\theta')\,.
\ee
We note the following useful integral over the Grassmann momenta
\beq
&& \int d^\NN p_\theta^\dagger e^{ \frac{p_\theta^{i\dagger} p_\theta^i}{\beta} } p_\theta^{i_1\dagger}  \ldots p_\theta^{i_q\dagger} = (-)^q \beta^{q-\NN} (\varepsilon p_\theta^{\NN-q})_{i_q\ldots i_1}\,,
\\
\label{06092019-man-18-ad01} &&  (\varepsilon p_\theta^{\NN-q})_{i_1\ldots i_q} \equiv \frac{1}{(\NN-q)!}  \varepsilon_{i_1\ldots i_q i_{q+1}\ldots i_\NN}  p_\theta^{i_{q+1}} \ldots p_\theta^{i_\NN}\,.
\eeq

Taking into account expression for $d\Gamma_\smp3^{p_\theta}$ obtained from \rf{03092019-man-09} by setting $n=3$ and notation in \rf{06092019-man-15}, we note the following helpful Berezin integrals for 3-point vertices:
\beq
\label{06092019-man-19} && \int d\Gamma_\smp3^{p_\theta^\dagger}  \exp\big(\sum_{a=1,2,3} \frac{ p_{\theta_a}^i  p_{\theta_a}^i\!\!{}^\dagger }{ \beta_a} \big) =  \beta^{-\NN} (\varepsilon \Pbf_\theta^\NN) (\varepsilon  \Po_\theta^\NN)\,, \hspace{1cm} \beta\equiv \beta_1\beta_2\beta_3\,,
\\
\label{06092019-man-20} && \int d\Gamma_\smp3^{p_\theta^\dagger}\,  (\varepsilon \Po_\theta^{\dagger\,\NN}) \exp\big(\sum_{a=1,2,3} \frac{ p_{\theta_a}^i  p_{\theta_a}^i{}\!\!^\dagger }{ \beta_a} \big) = (\varepsilon \Pbf_\theta^\NN)\,, \hspace{1.6cm} \Pbf_\theta^i \equiv \sum_{a=1,2,3}p_{\theta_a}^i\,, \qquad
\eeq
where $\Po_\theta^i\!{}^\dagger$ appearing in \rf{06092019-man-20} is obtained from \rf{04092019-man-04} by the replacement $p_{\theta_a}^i\rightarrow p_{\theta_a}^{i\dagger}$.

\appendix{ Superfield $\Phi_\lambda^*$  }

Using \rf{02092019-man-22} and notation in \rf{06092019-man-18-ad01}, we find the following expansion for the superfield $\Phi_\lambda^*$:
\be \label{07092019-man-01-ad01}
\Phi_\lambda^*(p,p_\theta) = \sum_{q=0}^{\NN} \frac{(-)^q}{q!}\beta^{\half q -\frac{1}{4}\NN + \half e_\lambda -\half e_{\lambda-\half q } } \phi_{\lambda - \half q + \frac{1}{4}\NN\,;\, i_1\ldots i_q }^\dagger(p) (\varepsilon  p_\theta^{\NN-q})_{i_1\ldots i_q}\,.
\ee
We note that equal-time (anti)commutator for the component fields \rf{02092019-man-09} takes the form
\be \label{07092019-man-01-ad02}
[\phi_{\lambda\,;\,i_1\ldots i_q}(p),\phi_{\lambda'\,;\,i_1'\ldots i_{q'}'}^\dagger(p')]_\pm = \frac{\beta^{-e_{ \lambda +\half }}}{2 (\NN-q)!}\delta^3(p-p') \delta_{\lambda\lambda'} \delta_{qq'} \varepsilon_{i_1\ldots i_q j_{q+1}\ldots j_\NN} \varepsilon^{i_1'\ldots i_q' j_{q+1}\ldots j_\NN}\,.
\ee
Using \rf{07092019-man-01-ad02}, we verify that (anti)commutator for superfields \rf{02092019-man-12},\rf{07092019-man-01-ad01} takes the form given in \rf{02092019-man-25}.

\noindent {\bf Realization of $\NN$-extended Poincar\'e superalgebra on superfield $\Phi_\lambda^*$}. Using \rf{02092019-man-14}-\rf{02092019-man-21} and \rf{02092019-man-23}, we get the realization of the $\NN$-extended Poincar\'e superalgebra on the superfield $\Phi_\lambda^*$ in terms of differential operators,
\beq
\label{07092019-man-01} && P^\Rsm = - p^\Rsm\,,  \qquad P^\Lsm = - p^\Lsm\,,   \hspace{1.4cm}    P^+ = - \beta\,,\qquad
P^- = -p^-\,, \quad p^- \equiv - \frac{p^\Rsm p^\Lsm}{\beta}\,,\qquad
\\
\label{07092019-man-02} && J^{+\Rsm}= \irm x^+ P^\Rsm + \partial_{p^\Lsm}\beta\,, \hspace{2.3cm} J^{+\Lsm}= \irm x^+ P^\Lsm + \partial_{p^\Rsm}\beta\,, \
\\
\label{07092019-man-03} && J^{+-} = \irm x^+P^- + \partial_\beta \beta + M_{-\lambda}^{+-}\,, \hspace{1cm} J^{\Rsm\Lsm} =  p^\Rsm\partial_{p^\Rsm} - p^\Lsm\partial_{p^\Lsm} + M_{-\lambda}^{\Rsm\Lsm}\,,
\\
\label{07092019-man-04} && J^{-\Rsm} = -\partial_\beta p^\Rsm + \partial_{p^\Lsm} p^-
+ M_{-\lambda}^{\Rsm\Lsm}\frac{p^\Rsm}{\beta} - M_{-\lambda}^{+-} \frac{p^\Rsm}{\beta}\,,
\\
\label{07092019-man-05} && J^{-\Lsm} = -\partial_\beta p^\Lsm + \partial_{p^\Rsm} p^-
- M_{-\lambda}^{\Rsm\Lsm}\frac{p^\Lsm}{\beta} - M_{-\lambda}^{+-} \frac{p^\Lsm}{\beta}\,,
\\
\label{07092019-man-06} && \hspace{1.2cm} M_\lambda^{+-} =  \half p_\theta^i \partial_{p_\theta^i} - \frac{1}{4} \NN - \half e_\lambda\,, \hspace{2cm} M_\lambda^{\Rsm\Lsm}  =   \lambda -\half p_\theta^i\partial_{p_\theta^i} + \frac{1}{4} \NN\,,
\\
\label{07092019-man-07} && Q_i^{+\Rsm} = (-)^{e_\lambda}\beta \partial_{p_\theta^i}\,, \hspace{2.8cm}  Q^{+\Lsm i} = (-)^{e_{\lambda+\half} } p_\theta^i\,,
\\
\label{07092019-man-08} && Q^{-\Rsm i} =    (-)^{e_{\lambda+\half} } \frac{1}{\beta} p^\Rsm p_\theta^i\,, \hspace{2cm}  Q_i^{-\Lsm} =  (-)^{e_\lambda} p^\Lsm \partial_{p_\theta^i}\,,
\\
\label{07092019-man-09} && J^i{}_j = p_\theta^i\partial_{p_\theta^j} - \frac{1}{\NN}\delta_j^i p_\theta^k \partial_{p_\theta^k}\,,
\eeq
where $e_\lambda$ is given in \rf{02092019-man-07}.

Use of relations in \rf{02092019-man-24},\rf{02092019-man-25}, leads to equal-time (anti)commutation relations between the generators of the Poincar\'e superalgebra and the superfield $\Phi_\lambda^*$,
\be \label{07092019-man-10}
[\Phi_\lambda^*,G_\smpt]_{\pm} =  G_{\diff,\,\lambda} \Phi_\lambda^* \,,
\ee
where $G_{\diff,\,\lambda}$ are given in \rf{07092019-man-01}-\rf{07092019-man-09}.

\noindent {\bf Hermitian conjugate of superfields and vertices}. Using \rf{02092019-man-22},\rf{02092019-man-23}, we verify that the hermitian conjugate of  $\Phi_\lambda^*$ can be presented as
\be \label{07092019-man-11}
(\Phi_\lambda^*(p,p_\theta))^\dagger \equiv \beta^{\frac{\NN}{2}}\int d^\NN p_\theta e^{ \frac{p_\theta^i p_\theta^{i\dagger} }{\beta} } \Phi_{-\lambda}^*(-p,p_\theta)\,.
\ee
The constraint on the coupling constants \rf{05092019-man-12} can be obtained in the following way. First, we introduce the vertices
\be  \label{07092019-man-12}
v^{\lambda_1\lambda_2\lambda_3} =  (\Po^\Lsm)^{\frac{\NN}{4}  + \Mbf_\lambda} \prod_{a=1,2,3} \beta_a^{-\lambda_a  -  \half e_{\lambda_a} }\,,
\hspace{0.5cm} \vb^{\lambda_1\lambda_2\lambda_3} = (\Po^\Rsm)^{\frac{\NN}{4}- \Mbf_\lambda}\, (\varepsilon\Po_\theta^\NN)  \prod_{a=1,2,3} \beta_a^{\lambda_a  - \half\NN -  \half e_{\lambda_a} }\,,
\ee
where $\Mbf_\lambda$ is given in  \rf{05092019-man-08}. Second, using \rf{07092019-man-11},\rf{07092019-man-12}, we get the relation
\be  \label{07092019-man-13}
\Big(\int d\Gamma_\smp3  \Phi_{\lambda_1\lambda_2\lambda_3}^*  v^{\lambda_1\lambda_2\lambda_3}\Big)^\dagger  = (-)^{\Mbf_\lambda } \int d\Gamma_\smp3 \Phi_{-\lambda_1,-\lambda_3,-\lambda_3}^* \vb^{-\lambda_1,-\lambda_2,-\lambda_3}\,.\qquad
\ee
Finally, with the help of \rf{07092019-man-13}, we see that, requiring the $P_\smp3^-$ to be hermitian, we obtain the constraint on coupling constants in \rf{05092019-man-12}.

\appendix{ Derivation of cubic vertex $p_{\lambda_1\lambda_2\lambda_3}^-$  \rf{05092019-man-01}.}

Our procedure of the derivation of cubic interaction vertex $p_\smp3^-$ given in \rf{05092019-man-01} is realized in the following five steps.

\noindent {\bf Step 1}. From \rf{04092019-man-37} we see that the vertex $p_\smp3^-$ can be presented as
\be \label{01062019-man02-01}
p_\smp3^- = V(\Po^\Lsm,\Po_\theta) + \Vb(\Po^\Rsm,\Po_\theta)\,.
\ee
$J^i{}_j$ symmetries \rf{04092019-man-34} imply that vertices $V$,$\Vb$ \rf{01062019-man02-01} can be presented as
\beq
\label{01062019-man02-02} && V(\Po^\Lsm,\Po_\theta) = V_0(\Po^\Lsm) + (\varepsilon \Po_\theta^\NN) V_\varepsilon(\Po^\Lsm)\,,
\\
&&  \label{01062019-man02-03} \Vb(\Po^\Rsm,\Po_\theta) = \Vb_0(\Po^\Rsm) + (\varepsilon \Po_\theta^\NN) \Vb_\varepsilon (\Po^\Rsm)\,,
\eeq
where we use notation in \rf{05092019-man-09} .

\noindent {\bf Step 2}. Using $p_\smp3^-$ \rf{01062019-man02-01} and requiring that the density $q_\smp3^{-\Lsm}$ \rf{04092019-man-35} be polynomial in the momentum $\Po^\Rsm$, we find $V_\varepsilon(\Po^\Lsm)=0$. Using $p_\smp3^-$ \rf{01062019-man02-01} and requiring that the density $q_\smp3^{-\Rsm}$ \rf{04092019-man-35} be polynomial in the momentum $\Po^\Lsm$, we find $\Vb_0(\Po^\Rsm)=0$.
We have the relations
\be  \label{01062019-man02-04}
V(\Po^\Lsm,\Po_\theta) = V_0(\Po^\Lsm)\,,  \qquad  \Vb(\Po^\Rsm,\Po_\theta) =  (\varepsilon \Po_\theta^\NN) \Vb_\varepsilon (\Po^\Rsm)\,.
\ee

\noindent {\bf Step 3}. Taking into account \rf{01062019-man02-04}, we learn that equations \rf{04092019-man-35},\rf{04092019-man-34} amount to the following
\beq
&& \hspace{-3cm} \hbox{ Equations for } \ V_0:
\nonumber\\[-5pt]
\label{01062019-man02-05} && \big(  N_{\Po^\Lsm} - \frac{\NN}{4} + \half \Ebf_\lambda + \sum_{a=1,2,3}\beta_a\partial_{\beta_a} \big) V_0 = 0\,,
\\
\label{01062019-man02-06} && \big( - N_{\Po^\Lsm} + \Mbf_\lambda  +\frac{\NN}{4}\big) V_0 = 0\,,
\\
\label{01062019-man02-07} && \big( - \No_\beta -  \Mo_\lambda   - \half \Eo_\lambda \big) V_0 =  0\,.
\\
&& \hspace{-3cm}  \hbox{ Equations for } \ \Vb_\varepsilon:
\nonumber\\[-5pt]
\label{01062019-man02-08} && \big( N_{\Po^\Rsm}+ \frac{5\NN}{4} + \half \Ebf_\lambda + \sum_{a=1,2,3}\beta_a\partial_{\beta_a} \big) \Vb_\varepsilon = 0\,,
\\
\label{01062019-man02-09} && \big( N_{\Po^\Rsm}    + \Mbf_\lambda  - \frac{\NN}{4} \big) \Vb_\varepsilon = 0\,,
\\
\label{01062019-man02-10} && \big( - \No_\beta +  \Mo_\lambda   + \half \Eo_\lambda \big) \Vb_\varepsilon =  0\,,
\eeq
where operators $N_{\Po^\Rsm}$, $N_{\Po^\Lsm}$ and $\No_\beta$  are given in \rf{04092019-man-11},\rf{04092019-man-30} and we use the notation
\be \label{01062019-man02-11}
\Mbf_\lambda \equiv \sum_{a=1,2,3} \lambda_a\,,\quad \Ebf_\lambda \equiv \sum_{a=1,2,3} e_{\lambda_a}\,, \quad  \Mo_\lambda = \frac{1}{3}\sum_{a=1,2,3}\betach_a \lambda_a\,,
\quad \Eo_\lambda = \frac{1}{3}\sum_{a=1,2,3}\betach_a e_{\lambda_a}\,.
\ee
Let us consider  the system of equations  \rf{01062019-man02-05}-\rf{01062019-man02-07}.

\noindent {\bf Step 4}. Equation \rf{01062019-man02-06} is solved as
\be \label{01062019-man02-12}
V_0 = (\Po^\Lsm)^{\Mbf_\lambda + \frac{\NN}{4} } V^{(1)}\,, \qquad V^{(1)} = V^{(1)}(\beta_1,\beta_2,\beta_3)\,,
\ee
where $V^{(1)}$ depends on the momenta $\beta_1,\beta_2,\beta_3$ and the helicities $\lambda_1,\lambda_2,\lambda_3$. Plugging \rf{01062019-man02-12} into  \rf{01062019-man02-05} and \rf{01062019-man02-07}, we obtain the respective equations
\beq
 \label{01062019-man02-13} &&  \big( \Mbf_\lambda + \half \Ebf_\lambda + \sum_{a=1,2,3}\beta_a\partial_{\beta_a} \big)V^{(1)} = 0\,,
\hspace{1cm} \big( \No_\beta +  \Mo_\lambda   + \half \Eo_\lambda \big) V^{(1)} =  0\,.\qquad
\eeq

\noindent {\bf Step 5}. Introducing new vertex $V^{(2)}$,
\be  \label{01062019-man02-15}
V^{(1)} = V^{(2)}  \prod_{a=1,2,3} \beta_a^{-\lambda_a - \half e_{\lambda_a} }\,,
\ee
we learn that equations \rf{01062019-man02-13} lead to the following two respective equations for  $V^{(2)}$:
\be  \label{01062019-man02-16}
\sum_{a=1,2,3}\beta_a\partial_{\beta_a} \, V^{(2)}   = 0\,, \qquad \No_\beta \, V^{(2)} =  0\,.
\ee
From \rf{01062019-man02-16}, we find that the $V^{(2)}$ does not dependent on the momenta $\beta_1$, $\beta_2$, $\beta_3$,
\be \label{01062019-man02-17}
V^{(2)} = C^{\lambda_1\lambda_2\lambda_3} \,,
\ee
where $C^{\lambda_1\lambda_2\lambda_3}$ is a constant depending only on the helicities.
Collecting formulas in \rf{01062019-man02-12}-\rf{01062019-man02-17},  we obtain vertex  $V_{\lambda_1\lambda_2\lambda_3}$ presented in \rf{05092019-man-02}. To determine the vertex $\Vb_\varepsilon$ we should analyse equations \rf{01062019-man02-08}-\rf{01062019-man02-10}. Repeating analysis above-given, we obtain solution to $\Vb_{\lambda_1\lambda_2\lambda_3}$ presented in \rf{05092019-man-03}.


\end{document}